\documentclass[prb,aps,twocolumn,reprint,longbibliography,floatfix]{revtex4-1}
\usepackage{amsmath,txfonts,graphicx,upgreek,xfrac}
\usepackage[bookmarks=true,colorlinks=true,urlcolor=blue,linkcolor=blue,citecolor=blue]{hyperref} 
\renewcommand{\Im}{\operatorname{Im}}
\renewcommand{\Re}{\operatorname{Re}}
\newcommand{\hev}{\hat{\textbf{e}}}
\newcommand{\kv}{\textbf{k}}
\newcommand{\rv}{\textbf{r}}
\newcommand{\qv}{\textbf{q}}
\newcommand{\dr}{d^3\rv}
\newcommand{\Deltax}{\Delta_{\textit{x}}^{\text{}}}
\newcommand{\DeltaKS}{\Delta_{\text{KS}}}
\newcommand{\LCAOTDDFTkomega}{LCAO-TDDFT-\emph{k}-$\omega$}
\newcommand{\GLLBsc}{GLLB-SC}
\newcommand{\Egap}{E_{\text{gap}}}
\begin{document}

\title{Optical Absorption and Energy Loss Spectroscopy of Single-Walled Carbon Nanotubes}

\author{Mar\'{\i}a Rosa Preciado-Rivas}
\author{Victor Alexander Torres-S\'{a}nchez}
\author{Duncan J. Mowbray}
\email{duncan.mowbray@gmail.com}
\affiliation{School of Physical Sciences and Nanotechnology, Yachay Tech University, Urcuqu\'{\i} 100119, Ecuador}

\begin{abstract}
  The recent development of efficient chirality sorting techniques has opened the way to the use of single-walled carbon nanotubes (SWCNTs) in a plethora of nanoelectronic, photovoltaic, and optoelectronic applications.  However, to understand the excitation processes undergone by SWCNTs, it is necessary to have highly efficient and accurate computational methods to describe their optical and electronic properties, methods which have until now been unavailable. Here we employ linear combinations of atomic orbitals (LCAOs) to represent the Kohn-Sham (KS) wavefunctions and perform highly efficient time dependent density functional theory (TDDFT) calculations in the frequency domain using our \LCAOTDDFTkomega{} code to model the optical absorbance and energy loss spectra and spatial distribution of the exciton charge densities in SWCNTs. By applying the \GLLBsc{} derivative discontinuity correction to the KS eigenenergies, we reproduce the measured $E_{11}$ and $E_{22}$ transitions within $\sigma \lesssim 70$~meV and the optical absorbance and electron energy loss spectra semi-quantitatively for a set of fifteen semiconducting and four metallic chirality sorted SWCNTs.  Furthermore, our calculated electron hole density difference $\Delta \rho(\rv, \omega)$ resolves the spatial distribution of the measured excitations in SWCNTs. These results open the path towards the computational design of optimized SWCNT nanoelectronic, photovoltaic, and optoelectronic devices \emph{in silico}.
\end{abstract}
\maketitle
 
\section{Introduction}

Single-walled carbon nanotubes (SWCNTs) have drawn attention in the field of organic electronics due to their unique physical properties, e.g., ballistic conductance, tailorable band gaps, photoluminescence, and high optical absorbance \cite{baughman2002carbon}. These nearly one dimensional (1D) structures come in various chiralities, which, depending on the way they are rolled up, change their energy band gaps yielding a plethora of different absorption and conductive properties \cite{Dresselhaus}. A variety of different semiconducting SWCNTs can be used to widen the range of wavelengths that can be potentially exploited in photovoltaic applications \cite{SpataruPRL2004, kataura1999optical}. SWCNTs exhibit intense absorption peaks with band gaps between 0.9 and 1.5~eV and have high thermal stability\cite{Yamamoto2008, Liew2005Thermal}. In the case of metallic nanotubes, electronic transport occurs ballistically, meaning they can carry high currents without heating \cite{liang2001fabry, frank1998carbon}. Furthermore, a clear advantage are the recently developed methods for separating SWCNTs based on their chirality. This provides a straightforward method for tailoring the band gap of the semiconducting layer in a solar cell.

For these reasons, SWCNTs have been widely used as additives in OPVs to improve their efficiency by increasing the charge carrier mobility of conventional polymers \cite{kymakis2002single, kymakis2006post} and dye-sensitized solar cells. In donor-acceptor hybrid cells, SWCNTs have been used to either covalently \cite{campidelli2008facile} or non-covalently \cite{bartelmess2010phthalocyanine} graft chromophore molecules, increasing incident photon to current efficiency (IPCE) by about 17\%. SWCNTs can interact with polymers via $\pi$-$\pi$ stacking, porphyrins electrostatically \cite{guldi2005single} to achieve an IPCE of 8.4\%, lipid nanodiscs, and human DNA \cite{Ham2010Photoelectrochemical}. Moreover, in many other photovoltaic devices, metallic carbon nanotubes are used as electrodes because of their ballistic conducting properties. % Baker 2013

Spectroscopy techniques are widely used to characterize SWCNTs. The advantages of optical absorbance (OA), a specific type of spectroscopy, rely on the fact that it is nondestructive, noninvasive, and simple to perform at room temperature and under ambient pressure. For instance, photoluminescence, absorption, and resonance Raman spectroscopy are widely employed in bulk SWCNTs samples in both research \cite{ZamoraLedezma2009Orientational, TorresCanas2014Dispersion} and industrial laboratories \cite{Weisman2019Introduction}. This makes spectroscopy techniques important for the development of OPVs as these methods provide insight into the properties of the materials, whether they are suitable for photovoltaic devices, and how they can be improved. For example, information about the exciton generation process can be gathered through spectroscopy techniques to make further improvements in the design of OPVs.  This is because, in the case of OA, light is most often absorbed when in resonance with the band gap of the material so that the observation of absorption peaks are related to electron transitions.

%For example, pump-probe spectroscopy has been used to study the excited state dynamics of the first optical transitions ${E}_{1 1}$ in SWCNTs \cite{LiviaMilan}. EELS has been used to experimentally study the optical properties and electronic transport correlated to the structure in carbon nanotubes \cite{SWCNTEELS}. On the other hand, experimental studies have given insight into the absorption spectrum and properties of both chlorophyll monomers and dimers with and without the proteins that contain them in biological structures \cite{Weisman2003Dependence, Stockett2015The, Milne2016On}. Moreover, experimental onset energies and absorption spectra of chlorophyll \textit{a} and \textit{b} have been reported to be blue-shifted by 30 to 70~nm compared to that of chlorophyll-containng proteins \cite{Milne2014Unraveling}.

Theoretical calculations of the photoabsorption processes in systems provide insight into not only how excitons are generated, but also other properties, such as the charge distribution, which can help to explain what is observed in experimental data. Some of the most commonly used methods are those based on density functional theory (DFT) \cite{Zangwill2015A}. DFT, based on the hypothesis that the electron density distribution completely characterizes the ground state of many electron systems, uses functionals of the spatially-dependent electron density to model the ground state electronic structure and properties at the quantum mechanical level. DFT has made important contributions in material design projects by combining theory and computational methods to replace traditional, and often expensive, experiments \cite{Nrskov2009Towards, Jain2013Commentary}. For instance, DFT calculations have been done to unravel the characteristics of spectroscopy for SWCNTs with linear response time-dependent density functional theory (TDDFT) used to complement the experimental work made in Ref.~\citenum{LiviaMilan}; additionally, estimates of the internal quantum efficiency of organic photovoltaic devices containing polymers, fullerene C$_{60}$ and SWCNTs have been obtained using DFT \cite{Glanzmann2016Theoretical}. 

\begin{figure*}
  \includegraphics[width=1.5\columnwidth]{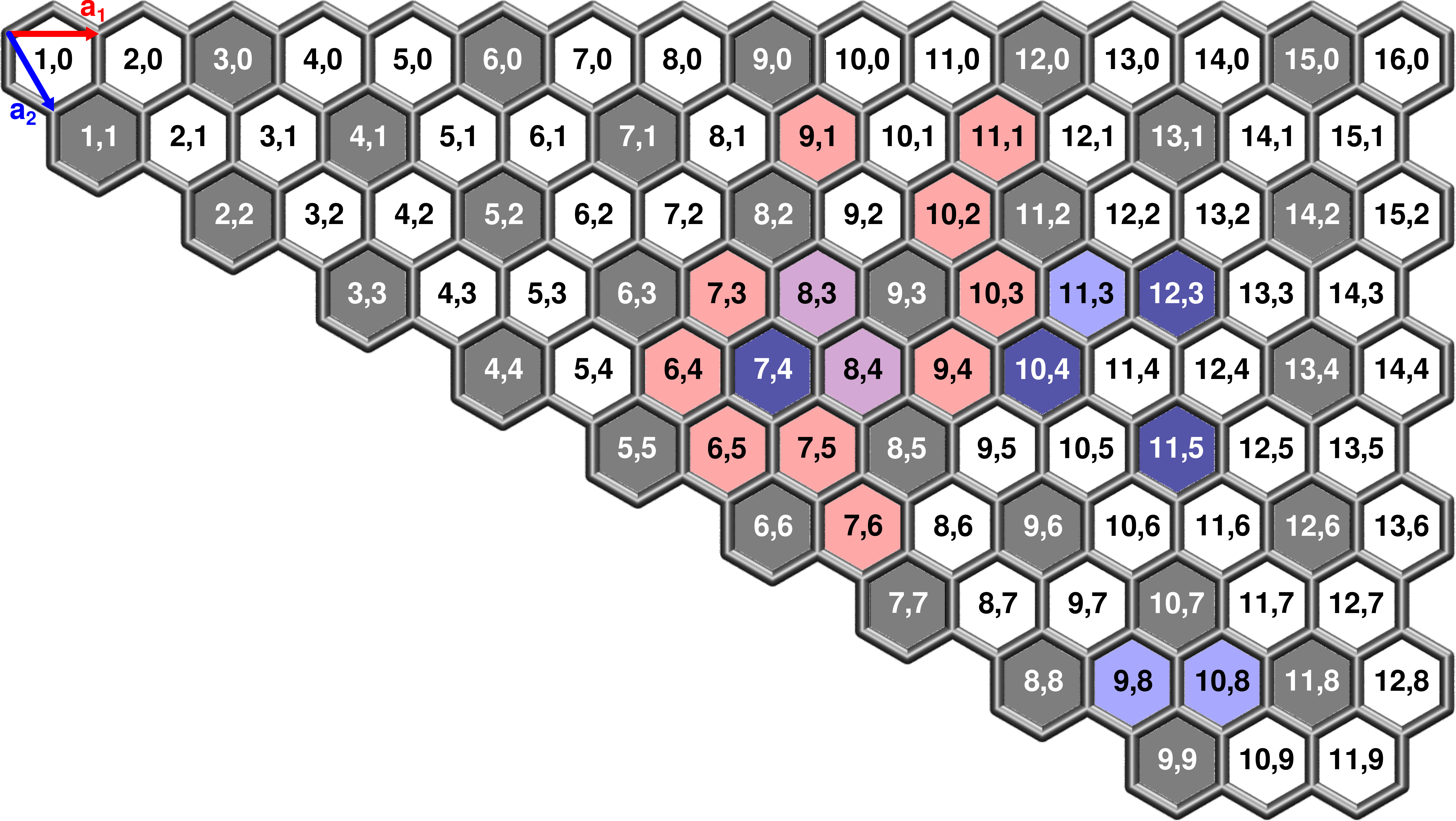}
  \caption{SWCNT indices $(m,n)$ of circumference vector $\textbf{C} \equiv m \textbf{a}_1 + n \textbf{a}_2$ where $\textbf{a}_1$ (red) and $\textbf{a}_2$ (blue) are the primitive unit vectors with optical absorbance (red), electron energy loss (blue), and both (mauve) data from Refs.~\citenum{OpticalAbsorbance} and \citenum{SWCNTEELS}, respectively. Metallic tubes ($m-n = 0 \textrm{ mod }3$) are marked in grey or dark blue.}\label{fig:FigSWNTmap}
\end{figure*}

Optical selection rules for SWCNTs allow light polarized parallel to the nanotube's axis to excite intense transitions between the corresponding subbands in the valence and conduction bands. For instance, v$_1$$\to$c$_1$ and v$_2$$\to$c$_2$, and so on, correspond to well-defined absorption transitions between van Hove singularities with energies $E_{11}$ and $E_{22}$. Metallic SWCNTs also have intense absorption peaks associated to transitions between van Hove singularities \cite{Weisman2006vanHoveE}.  Recently, experimental measurements of chirality sorted SWCNTs have provided both optical absorbance \cite{OpticalAbsorbance} and electron energy loss  spectra \cite{SWCNTEELS} for a large variety of SWCNTs.  For this reason, SWCNTs provide experimentally relevant 1D periodic systems for benchmarking our \LCAOTDDFTkomega{} code\cite{LiviaMilan,LCAOTDDFTKeenan,Chlorophyll,LCAOTDDFTkomega}.

In this work, we employ linear combinations of atomic orbitals (LCAO) to represent the Kohn-Sham (KS) wavefunctions within time dependent density functional theory (TDDFT) in momentum $\kv$ and frequency $\omega$ space, using our \LCAOTDDFTkomega{} code\cite{LiviaMilan,LCAOTDDFTKeenan,Chlorophyll,LCAOTDDFTkomega}, applying the derivative discontinuity correction\cite{GLLBSC}  to the KS eigenenergies.  This method is applied to the set of four metallic and fifteen semiconducting SWCNTs mapped in Figure~\ref{fig:FigSWNTmap}, for which optical absorbance and electron energy loss  spectroscopy measurements of chirality sorted samples are available from Refs.~\citenum{OpticalAbsorbance} and \citenum{SWCNTEELS}, respectively. Employing the exciton density method implemented within our \LCAOTDDFTkomega{} code\cite{LCAOTDDFTKeenan,LCAOTDDFTkomega}, we are able to provide a spatially resolved description of the experimentally observed transitions in metallic and semiconducting SWCNTs.

The paper is organized as follows. In Sec.~\ref{Methodology} we begin by providing a brief theoretical background in Sec.~\ref{TheoreticalBackground} to the derivative discontinuity correction obtained from the exchange part of the \GLLBsc{} functional $\Deltax$, the \LCAOTDDFTkomega{} method in the optical limit $\|\qv\|\rightarrow 0^+$,  and our model for the spatial distribution of the exciton charge density $\Delta\rho = \rho_e + \rho_h$, followed by a complete description of the relevant parameters employed in our DFT and \LCAOTDDFTkomega{} calculations in Sec.~\ref{ComputationalDetails}.  In Sec.~\ref{ResultsDiscussion} we compare our results with those obtained from experiments for the nineteen SWCNTs studied (see Figure~\ref{fig:FigSWNTmap}), including the atomic and electronic structure in Sec.~\ref{BandGaps}, optical absorbance spectra in Sec.~\ref{sec:SWCNTOpticalAbsorption}, the $E_{11}$ and $E_{22}$ transitions in semiconducting SWCNT in Sec.~\ref{E11E22}, and electron energy loss spectroscopy  in Sec.~\ref{EELS}, followed by our spatially resolved description of the exciton density for the $E_{11}$ transition in semi-conducting SWCNTs and the Dirac plasmon in metallic SWCNTs in Sec.~\ref{ExcitonDensity}.  Finally, concluding remarks are provided in Sec.~\ref{Conclusions}.  Atomic units ($\hslash = e = m_e = a_0 = 1$) have been employed throughout unless otherwise noted.

\section{Methodology}\label{Methodology}
\subsection{Theoretical Background}\label{TheoreticalBackground}

%\subsubsection{\LCAOTDDFTkomega{}}

%\subsubsection{Derivative Discontinuity Correction}\label{sec:GLLBSC}

Modelling the optical absorbance or electron energy loss  spectra of a material requires a proper description of its electronic structure, including the electronic band gap $\Egap$.  At the Kohn-Sham (KS) level the band gap is approximated by the energy difference between the KS eigenenergies, $\DeltaKS  = \varepsilon_{N+1} - \varepsilon_N$, where $N$ is the number of electrons and we have suppressed dependence on spin and $k$-point.  However, $\DeltaKS$ often underestimates the experimental band gap by an order of magnitude.  Although the exchange and correlation (xc) potential can be tuned to obtain a better agreement of $\Delta_\text{KS}$ with $\Egap$, this can lead to a potential that has unphysical features, resulting in a poor description of properties other than the band gap \cite{Tran2017Importance}.  While both hybrid functionals (HSE06\cite{HSE}) and quasiparticle methods ($G_0 W_0$\cite{AngelGWReview,OurJACS}) often provide a sufficiently accurate description of the electronic structure, their intractability makes such methods unsuitable for large macromolecules such as the SWCNTs we will study herein.

The derivative discontinuity correction to the exchange functional, $\Deltax$, has been proposed as a first-order \emph{ab initio} correction to the KS band gap\cite{Tran2018GLLBsc3}, where $\Egap \approx \DeltaKS + \Deltax$.  Kuisma \emph{et al.}\cite{GLLBSC} calculated the exchange part of the derivative discontinuity $\Delta_{\textit{x}}$ from the KS equations by using a modified version of the Gritsenko, van Leeuwen, van Lenthe, and Baerends (GLLB) xc potential \cite{Gritsenko1995GLLB, Gritsenko1997GLLB2}. This xc potential exhibits a step structure at the lowest unoccupied orbital when it starts to be occupied.

A newer version of this potential is called \GLLBsc{}, for solid and correlation, and has been shown to yield a better agreement with the experimental band gaps than LDA or GGA for solids \cite{Castelli2012GLLBsc2}.  The derivative discontinuity correction of the exchange part of the \GLLBsc{} functional is given by
\begin{equation}
  \Deltax = \frac{8\sqrt{2}}{3\pi^2} \sum_{n=1}^N \left(\sqrt{\varepsilon_{N+1} - \varepsilon_n} - \sqrt{\varepsilon_{N} - \varepsilon_n}\right)\langle\psi_{N+1}|\frac{\psi_{n}^*\psi_{n}^{}}{\rho}|\psi_{N+1}\rangle,\label{eq:Deltax}
\end{equation}
where  $N$ is the number of electrons, $\psi_n$ and $\varepsilon_n$ are the $n$th Kohn-Sham (KS) wavefunction and eigenenergy, respectively, and we have suppressed dependence on the spin and $k$-point.  

Major advantages of employing $\Deltax$ are both its \emph{ab initio} nature and its efficiency.  Specifically, the calculation of $\Deltax$ requires a single-point calculation of the electronic structure for the relaxed geometry, and the summation given in Eq.~\ref{eq:Deltax}.  This makes the derivative discontinuity correction an attractive alternative to hybrid functionals or quasiparticle methods for accurately describing the electronic structure of large macromolecules.

%%\subsubsection{Optical Absorbance}
%\subsubsection{TDDFT in $(k,\omega)$ Space and the Optical Limit}
%We begin by...
We model the optical absorption and electron energy loss  spectra using the head of the dielectric function, $\varepsilon(\omega)$, from our \LCAOTDDFTkomega{} code\cite{LiviaMilan,LCAOTDDFTKeenan,Chlorophyll,LCAOTDDFTkomega}, neglecting local crystal field effects. Adding the derivative discontinuity correction of the exchange part of the \GLLBsc{} functional $\Deltax$ from Eq.~\ref{eq:Deltax} to the eigenenergies of unoccupied KS states, the dielectric function is then\cite{LiviaMilan,AngelGWReview}
%\begin{widetext}
\begin{equation}
  %  \varepsilon_{\text{M}}(\omega) = 1 - \frac{4\pi}{\Omega} \sum_{\kv}\sum_{nm} \frac{w_\kv(f(\varepsilon_{\kv,m}) - f(\varepsilon_{\kv,n}))}{\omega - \varepsilon_{\kv,n} + \varepsilon_{\kv,m} - \Deltax + i\eta} \left|\frac{\hev_{\qv}\cdot\langle\psi_{\kv,n}|{\mathbf{\nabla}} |\psi_{\kv,m}\rangle}{\varepsilon_{\kv,n} - \varepsilon_{\kv,m} + \Deltax}\right|^2. \label{eq:epsilon}
    \varepsilon(\omega) = 1 - \frac{4\pi}{\Omega} \sum_{nm} \frac{f(\varepsilon_{m}) - f(\varepsilon_{n})}{\omega - (\varepsilon_{n} - \varepsilon_{m} + \Deltax) + i\eta} \left|\frac{\hev_{\qv}\cdot\langle\psi_{n}|{\mathbf{\nabla}} |\psi_{m}\rangle}{\varepsilon_{n} - \varepsilon_{m} + \Deltax}\right|^2, \label{eq:epsilon}
\end{equation}
%\end{widetext}
where we have suppressed spin and $k$-point dependence, $f$ is the Fermi-Dirac function, $\eta \approx 25$~meV is the Lorentzian broadening of the peaks, and $\hev_{\qv}$ is a unit vector in the direction of the light's polarization $\qv \rightarrow 0^+$.  

The matrix elements in Eq.~\ref{eq:epsilon} are expressed using the PAW formalism as
\begin{equation}
  \langle\psi_{n}|{\mathbf{\nabla}} |\psi_{m}\rangle = \sum_{a,a'}\sum_{i j} c_{in}^{a\dagger}c_{j m}^{a'} \langle \tilde{\phi}_{i}^a | \mathcal{T}^\dagger \nabla \mathcal{T} | \tilde{\phi}_{j}^{a'}\rangle,\label{eq:MatrixElements}
\end{equation}
where $\tilde{\phi}_{i}^a$ is the $i$th smooth basis function centered on atom $a$ and $\mathcal{T}$ is the PAW transformation operator
\begin{equation}
  \mathcal{T} = 1 + \sum_{a}\sum_{i} \left(|\varphi_i^a\rangle - |\tilde{\varphi}_i^a\rangle\right)\langle \tilde{p}_{i}^a|,
\end{equation}
where $\varphi_i^a$ and $\tilde{\varphi}_i^a$ are the all-electron and pseudo partial waves for state $i$ on atom $a$ and $\tilde{p}_{i}^a$ are their smooth PAW projector functions.

Methods for calculating Eq.~\ref{eq:MatrixElements} are already implemented within DFT to obtain the forces, i.e., the expectation value of the gradient operator within the LCAO basis. For this reason, obtaining the dielectric function $\varepsilon(\omega)$ within the \LCAOTDDFTkomega{} code simply involves the multiplication of matrices that have already been calculated, i.e., the KS coefficient matrices $c_{in}^a$ with the expectation values of the gradient operator in the PAW-corrected LCAO basis $\langle \tilde{\phi}_{i}^a|\mathcal{T}^{\dagger}\nabla \mathcal{T}|\tilde{\phi}_{j}^{a'}\rangle$.

This is a very efficient method with a scaling better than $\mathcal{O}(NM^2)$ where $N$ is the number of KS wavefunctions and $M \geq N$ is the total number of basis functions used in the LCAO calculation.  Moreover, the implicit summation over spin, $k$-point, and domain in Eq.~\ref{eq:epsilon}  lends itself trivially to parallelization employing the facilities available within most DFT codes.  This degree of parallelizability, as implemented within the \LCAOTDDFTkomega{} code\cite{LCAOTDDFTKeenan,LCAOTDDFTkomega}, proved essential for performing distributed memory calculations of SWCNTs with large unit cells ($\sim 50$~\AA{}) employing the dense $k$-point sampling ($\Delta k \lesssim \frac{1}{1200}$~nm$^{-1}$) required to converge the room temperature ($\eta\approx 25$~meV) optical absorbance.

It should be noted that employing an LCAO representation to solve for the non-zero dielectric matrix elements, outside the optical limit $\qv\rightarrow 0^+$, is unfeasible.  This is because the LCAO representation does not lend itself to the efficient calculation of Fourier transforms, unlike real-space and plane-wave methods.  For this reason, the \LCAOTDDFTkomega{} code's range of applicability is restricted to the optical limit with local crystal field effects neglected.  However, as we shall see, this simplification, when the \GLLBsc{} derivative discontinuity correction is employed, leads to a semi-quantitative description of optical and energy loss spectra for SWCNTs.

%\subsubsection{Exciton Charge Density}

We model the exciton density as the electron hole density difference, $\Delta \rho(\rv,\omega) = \rho_{\text{h}}(\rv,\omega)+\rho_{\text{e}}(\rv,\omega)$,  where the electron/hole densities are obtained by averaging over the hole/electron position, as implemented in our \LCAOTDDFTkomega{} code\cite{LCAOTDDFTKeenan,LCAOTDDFTkomega}.  This may be calculated using \cite{Livia2014PSSB,CatecholExcitons}
\begin{equation}
  %\Delta\rho(\rv,\omega) \approx  \sum_{\kv}\sum_{nm} \frac{\eta^2 w_{\kv} |f_{\kv,nm}|^2 \left(|\psi_{\kv,m}(\rv)|^2 - |\psi_{\kv,n}(\rv)|^2\right)}{(\omega-\varepsilon_{\kv,n}+\varepsilon_{\kv,m} - \Deltax)^2 + \eta^2}, \label{eq:exciton_difference}
  \Delta\rho(\rv,\omega) \approx  \sum_{nm} \frac{\eta^2 |\tau_{m\to n}|^2 \left(|\psi_{m}(\rv)|^2 - |\psi_{n}(\rv)|^2\right)}{(\omega-(\varepsilon_{n}-\varepsilon_{m} + \Deltax))^2 + \eta^2}, \label{eq:exciton_difference}
\end{equation}
where we have suppressed spin and $k$-point dependence and $\int\Delta\rho(\rv,\omega) \dr =0$.  Here $|\tau_{m\to n}|^2$ are the calculated intensities of the $m \to n$ transition from $\Im[\varepsilon(\omega)]$ of Eq.~\ref{eq:epsilon}, so that
\begin{equation}
  \Im[\varepsilon(\omega)] = \int \rho_h(\rv,\omega)\dr= -\int \rho_e(\rv,\omega)\dr. 
\end{equation}
In this way we take into account the relative strength of transitions and their contribution at a given frequency $\omega$.

\subsection{Computational Details}\label{ComputationalDetails}

In Figure~\ref{fig:FigSWNTmap} we show the indices $(m,n)$ of the SWCNTs for which we have performed calculations. Those marked in red are semiconducting SWCNTs with optical absorption spectra, in blue are semiconducting SWCNTs with electron energy loss spectra, in mauve are those with both optical absorption and electron energy loss spectra, and those in dark blue are metallic SWCNTs with electron energy loss spectra, as taken from Refs.~\citenum{OpticalAbsorbance} and \citenum{SWCNTEELS}.

%\subsection{Computational details}nnnnnnn
Our density functional theory (DFT) calculations were performed using the \textsc{gpaw} code\cite{GPAW,GPAWrev}, based on the projector-augmented wave (PAW) method\cite{PAW,GPAW} within the atomic simulation environment (ASE)\cite{ASE0,ASE}. We have used for the SWCNTs a revised Perdew-Burke-Ernzerhof generalized gradient approximation (GGA) for solids (PBEsol) \cite{PBEsol} for the exchange and correlation (xc) functional, and represented the Kohn-Sham (KS) wave functions using a linear combination of atomic orbitals (LCAO)\cite{GPAWLCAO} with a double-$\zeta$-polarized (DZP) basis set.  A room temperature electronic broadening of $\eta = k_B T = 25$~meV was employed throughout.

Both the unit cell and atomic structure for each of the nineteen SWCNTs studied (see Figure~\ref{fig:FigSWNTmap}) were relaxed until the maximum force was less than 0.05 eV/\AA{} by including 10~\AA{} of vacuum perpendicular to the SWCNT's axis. Periodic boundary conditions were employed only in the direction of the SWCNT axis, with the electron density and KS wave functions set to zero at the unit cell boundaries perpendicular to the SWCNT's axis. A grid spacing of $h \approx 0.2$~\AA{}  was employed and the Brillouin-zone was sampled with a $k$-point density of $\Delta k \lesssim$~$\frac{1}{30}$~\AA{}$^{-1}$ along the SWCNT's axis.

A Harris calculation was performed for each SWCNT to increase the $k$-point density to $\Delta k \lesssim \frac{1}{1200}$~nm$^{-1}$, fixing the electron density throughout the self-consistency cycle.  Such a dense $k$-point density was found to be necessary to converge the calculated absorbance spectra at room temperature ($\eta = k_B T = 25$~meV). In order to improve the description of the electronic gap, we employed the derivative discontinuity correction to the exchange part of the \GLLBsc{} functional $\Deltax$ as provided in Eq.~\ref{eq:Deltax}, by performing a single-point calculation for the relaxed structures with \GLLBsc{}.

All calculations of the dielectric function $\varepsilon(\omega)$ and electron hole density different $\Delta\rho(\rv,\omega)$ were performed using linear combinations of atomic orbitals (LCAOs) to represent the KS wave functions at the time-dependent density functional theory (TDDFT) level in the optical limit ($\qv \to 0^+$)  in reciprocal $k$-space and the frequency $\omega$ domain using our \LCAOTDDFTkomega{} code\cite{LiviaMilan,LCAOTDDFTKeenan,Chlorophyll,LCAOTDDFTkomega}.  Here we have employed a room temperature Lorentzian broadening ($\eta = 25$~meV) to the peaks and corrected the eigenenergies by the derivative discontinuity correction $\Deltax$ from Eq.~\ref{eq:Deltax} when calculating the dielectric function using Eq.~\ref{eq:epsilon}. We model the optical absorbance spectra using $\Im[\varepsilon(\omega)]$  and the electron energy loss  spectra using $-\Im[\varepsilon^{-1}(\omega)]$\cite{AngelGWReview}.% = \frac{\Im[\varepsilon(\omega)]}{\Re[\varepsilon(\omega)]^2 + \Im[\varepsilon(\omega)]^2}$. 

\section{Results \& Discussion}\label{ResultsDiscussion}

\subsection{Atomic and Electronic Structure}\label{BandGaps}
\begin{table}
  \caption{Relaxed single-walled carbon nanotube (SWCNT) diameters $d$ in \AA{}, unit cell lengths $L$ in \AA{}, numbers of atoms $N_{\text{at}}$ per unit cell, and derivative discontinuity corrections $\Deltax$ and electronic band gaps $\Egap$ in eV.}\label{tab:SWCNTs}
\begin{tabular}{c@{\qquad}r@{.}l@{\qquad}r@{.}l@{\qquad}r@{\qquad}r@{.}l@{\quad}r@{.}l}
  \hline\hline
SWCNT & \multicolumn{2}{c@{\qquad}}{$d$} & \multicolumn{2}{c@{\qquad}}{$L$} & $N_{\text{at}}$ & \multicolumn{2}{c@{\qquad}}{$\Deltax$} & \multicolumn{2}{c}{$\Egap$}\\
&  \multicolumn{2}{c@{\qquad}}{(\AA{})} & \multicolumn{2}{c@{\qquad}}{(\AA{})} & & \multicolumn{2}{c@{\qquad}}{(eV)} & \multicolumn{2}{c}{(eV)}\\
\hline
    (6,4) & 6&97 & 18&64 & 152 & 0&418 &1&475\\
    (9,1) & 7&62 & 40&81 & 364 & 0&417 &1&476\\
    (8,3) & 7&84 & 42&12 & 388 & 0&399 &1&414\\
    (6,5) & 7&60 & 40&83 & 364 & 0&374 &1&317\\
    (7,3) & 7&09 & 38&06 & 316 & 0&365 &1&281\\
    (7,5) & 8&31 & 44&68 & 436 & 0&360 &1&273\\
    (10,2)& 8&84 & 23&80 & 248 & 0&358 &1&269\\
    (9,4) & 9&17 & 49&33 & 532 & 0&339 &1&199\\
    (8,4) & 8&40 & 11&34 & 112 & 0&329 &1&158\\
    (7,6) & 8&97 & 48&21 & 508 & 0&315 &1&113\\
    (10,3)& 9&36 & 50&47 & 556 & 0&278 &0&983\\
    (11,1)& 9&16 & 49&35 & 532 & 0&269 &0&950\\

    (10,8) & 12&38 & 33&37 & 488 & 0&240 &0&849\\
    (9,8) & 11&66 & 63&02 & 868 & 0&246 &0&869\\
    (11,3) & 10&13 & 54&63 & 652 & 0&310 &1&098\\

    (11,5)& 11&22 & 20&20 & 268 & \multicolumn{2}{c@{\qquad}}{---} & \multicolumn{2}{c}{---}\\
    (12,3)& 10&89 & 6&51 & 84 & \multicolumn{2}{c@{\qquad}}{---} & \multicolumn{2}{c}{---}\\
    (10,4)& 9&91 & 8&91 & 104 & \multicolumn{2}{c@{\qquad}}{---} & \multicolumn{2}{c}{---}\\
    (7,4) & 7&70 & 13&69 & 124 & \multicolumn{2}{c@{\qquad}}{---} & \multicolumn{2}{c}{---}\\
%    (10,0) & 7&96 & 4&30 & 40 \\
%    (10,10) & 13&71 & 2&46 & 40 \\
\hline\hline
\end{tabular}
\end{table}

We will begin our analysis of the SWCNTs shown schematically in Figure~\ref{fig:FigSWNTmap} by considering their atomic and electronic structure.  As shown in Table~\ref{tab:SWCNTs}, this is a rather diverse selection of SWCNTs, with diameters ranging from $6.97$ to $12.38$ \AA, lengths from $6.51$ to $63.02$ \AA, and from $84$ to $868$ atoms per unit cell. Moreover, they exhibit different electronic properties, with four of them being metallic SWCNTs and the remaining fifteen semiconducting SWCNTs with band gaps between 0.8 and 1.5~eV.
%they are either metallic or semiconducting according to their ($m$,$n$) indices.

In Table \ref{tab:SWCNTs} we also provide the derivative discontinuity correction of the exchange part of the \GLLBsc{} functional\cite{GLLBSC}, $\Deltax$,  calculated using Eq.~\ref{eq:Deltax}. These corrections have a size on average of $\sim 28\%$ of the corrected band gap ($ \Deltax \approx 0.28 \Egap$), and are thus proportional to both the corrected band gap energy $\Egap$ and the KS band gap $\DeltaKS$.  More specifically, for the SWCNTs considered herein, $\Deltax$ ranges from 0.24 to 0.42~eV, as shown in Table~\ref{tab:SWCNTs}.  This implies $\Deltax$ will provide a qualitative correction to both the onset and intensities of the calculated spectra.  As we will see in the following sections, this correction is essential for providing both a semi-quantitative and qualitative description of optical absorption and electron energy loss spectra in the $\qv \to 0^+$ limit.

\subsection{Optical Absorption Spectra}\label{sec:SWCNTOpticalAbsorption}

\begin{figure}[!t]
  \includegraphics[width=\columnwidth]{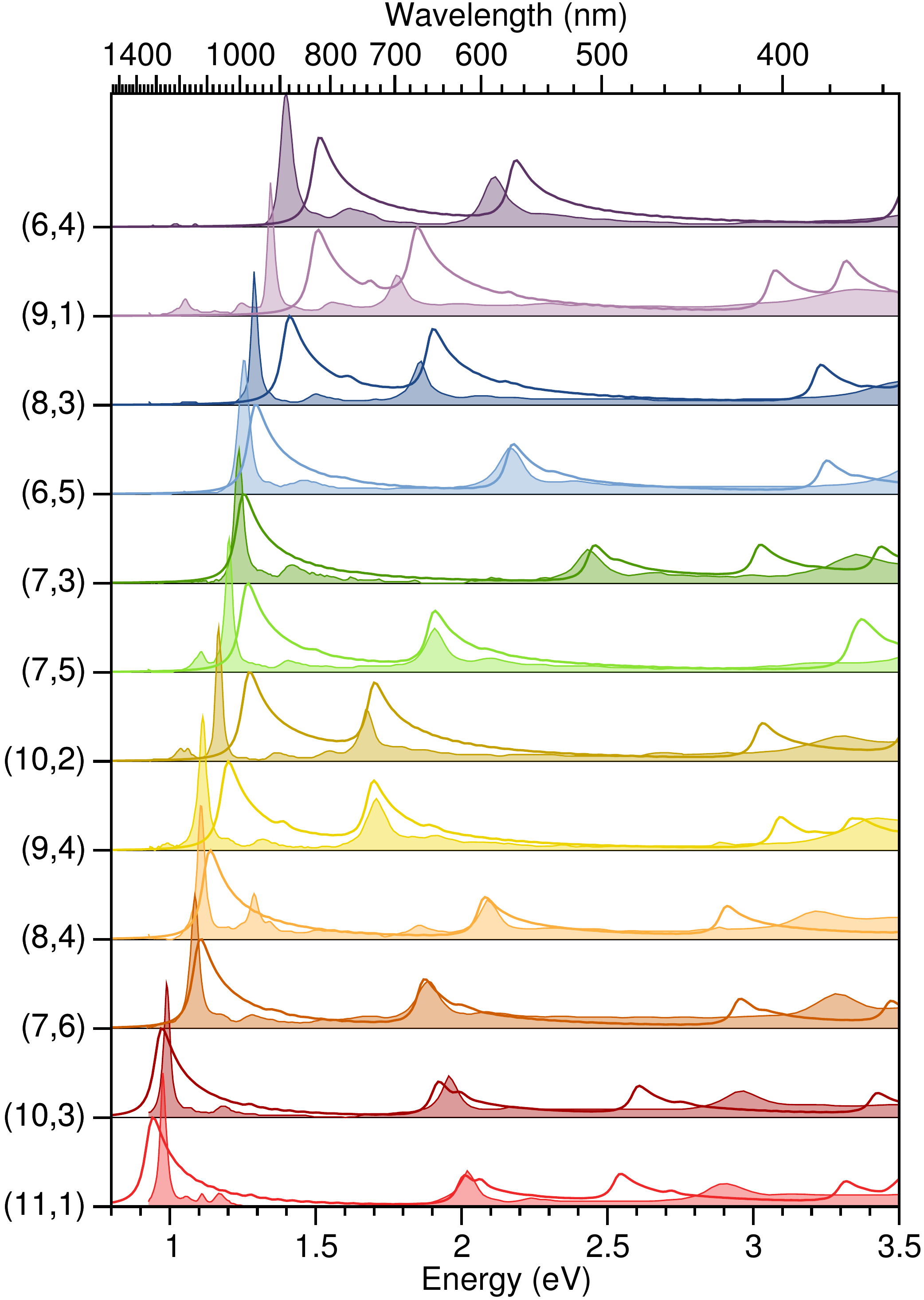}
  \caption{Comparison of \LCAOTDDFTkomega{} calculated (solid lines) and measured (filled regions, Ref.~\citenum{OpticalAbsorbance}) optical absorbance $\Im[\varepsilon(\omega)]$ spectra along the SWCNT axis in nm (upper axis) and eV (lower axis) for chirality sorted (6,4), (9,1), (8,3), (6,5), (7,3), (7,5), (10,2), (9,4), (8,4), (7,6), (10,3), and (11,1) SWCNTs shown in Figure~\ref{fig:FigSWNTmap}.}\label{fig:SWCNTOpticalAbsorbance}
\end{figure}

In Figure~\ref{fig:SWCNTOpticalAbsorbance}, we directly compare the optical absorption spectra calculated with our \LCAOTDDFTkomega{} code to the experimental data provided in Ref.~\citenum{OpticalAbsorbance}. The experimental data was normalized, that is, the highest value was set to 1.5 in arbitrary units. Likewise, we normalized the maximum of the calculated spectra to 1 in the same arbitrary units.

In each of the optical absorption spectra, the first peak corresponds to the first excitation associated to van Hove singularities, i.e., the  $E_{11}$ transition. The second highest peak corresponds to the second excitation of this kind, $E_{22}$. We were able to resolve in some cases a peak between these two, such as in the (9,4), (9,1), (8,3), (7,5) and (9,4). We can also observe excitations higher in energy than the $E_{22}$, which are typically red shifted by about 0.3~eV relative to the experimental peaks. From hereon, we shall restrict our discussion to the $E_{11}$ and the $E_{22}$ excitations.

All spectra were calculated using the derivative discontinuity correction of the exchange part of the \GLLBsc{} functional $\Deltax$ from Eq.~\ref{eq:Deltax}. As seen in Eq.~\ref{eq:epsilon}, this not only shifts the KS eigenenergies, yielding a good description of the first excitation energy $E_{11}$ or the band gap $\Egap$, but also changes the intensities of all the transitions.

For all the SWCNTs studied herein, we obtained a semi-quantitative agreement in the description of the relative intensities and positions of the $E_{11}$ and $E_{22}$ transitions. We find that the $E_{11}$ transition is more intense than the $E_{22}$, except in the case of the (9,1) SWCNT. The first transition being more intense than the second transition is a feature also observed in the experimental data.

When comparing the spectra of all the SWCNTs, we find that the energies of the $E_{22}$ transitions are uncorrelated to those of the $E_{11}$ transitions. The energies of these two transitions are neither separated by the same amount nor is one proportional to the other. In some cases they are closer than in others. Although the $E_{11}$ and the $E_{22}$ transitions are uncorrelated, the spectra calculated with the \LCAOTDDFTkomega{} code was able to reproduce semi-quantitatively the values of both these transitions energies.

The optical absorption spectra shown in Figure~\ref{fig:SWCNTOpticalAbsorbance} corresponds to the first 12 SWCNTs listed in Table \ref{tab:SWCNTs}, which vary widely in length and number of atoms per unit cell. We can observe that the spectra calculated using the \LCAOTDDFTkomega{} code reproduce the features of the experimental data of these very different SWCNTs, showing that it is surprisingly robust when calculating the optical absorbance of carbon-based 1D nanostructures.

In order to calculate the optical absorption spectra of SWCNTs with large unit cells, it was also necessary to implement both domain decomposition of the real space grids and parallelization with respect to $k$-points. This type of parallelization proved essential for allowing us to perform distributed memory calculations with limited computational resources.

\subsection{\emph{E}$_{\text{11}}$ and \emph{E}$_{\text{22}}$ Transitions}\label{E11E22}

\begin{figure*}[!t]
  \includegraphics*[width=2\columnwidth]{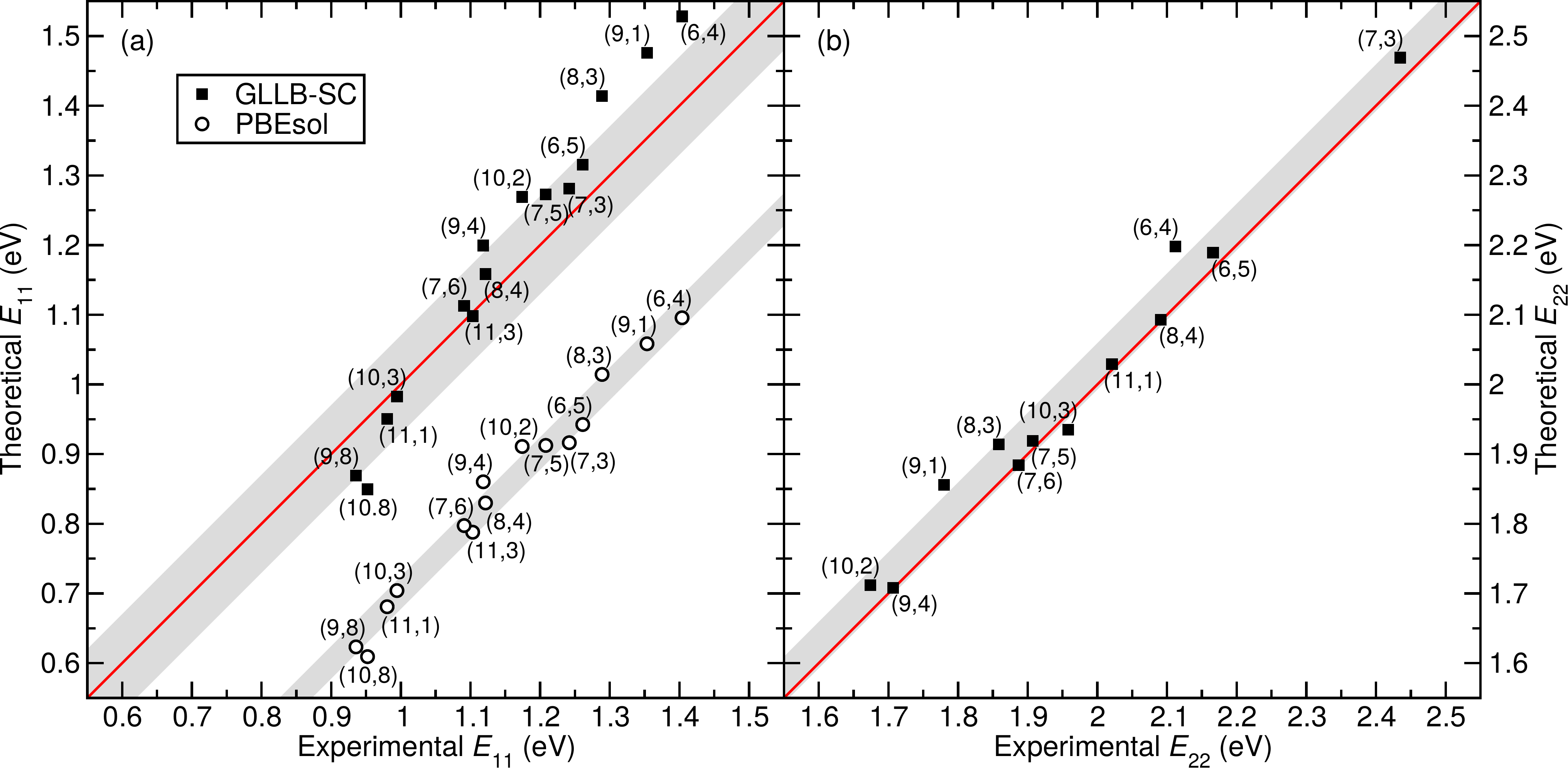}
  \caption{Theoretical versus experimental (a) $E_{11}$ and (b) $E_{22}$ transition energies in eV from \LCAOTDDFTkomega{} including (\GLLBsc{}\cite{GLLBSC}, filled squares) and neglecting (PBEsol\cite{PBEsol}, open circles) the derivative discontinuity correction $\Deltax$ and from optical absorbance and electron energy loss measurements of Refs.~\citenum{OpticalAbsorbance} and \citenum{SWCNTEELS}, respectively, for 15 different SWCNT chiralities.  The average errors for \GLLBsc{} (a) $E_{11}$ ($\epsilon \approx 0\pm 70$~meV) and (b) $E_{22}$ ($\epsilon \approx 20 \pm 33$~meV) transitions and for PBEsol (a) $E_{11}$ ($\epsilon \approx -300\pm3$~meV) transitions are shown as grey regions. Red lines are provided to guide the eye.
  }
  \label{fig:E11E22}
\end{figure*}

In Figure~\ref{fig:E11E22}(a) we directly compare the measured $E_{11}$ transition energies from optical absorption and electron loss spectroscopy of Refs.~\citenum{OpticalAbsorbance} and \citenum{SWCNTEELS}, respectively, with our \LCAOTDDFTkomega{} calculated values. We also compare the values obtained with the derivative discontinuity correction of the exchange part of the \GLLBsc{} functional $\Deltax$ and without this correction, that is, using only the PBEsol xc functional.

The PBEsol functional yields an estimation of the band gap with an average error of $\epsilon \approx -0.30 \pm 0.02$~eV. Although it reproduces the trend better than in the case of \GLLBsc{} functional, the band gap is always significantly underestimated. When we add the derivative discontinuity correction of the \GLLBsc{} functional, the average error is $\epsilon \approx 0 \pm 0.07$~eV, that is, much smaller than when not adding the correction and well within the expected 0.1 eV accuracy of DFT calculations. Nevertheless, the standard deviation is somewhat larger. This shows that it is important to use the derivative discontinuity correction of the \GLLBsc{} functional to properly describe the electronic structure and have a better agreement with the experimentally measured spectra onset for SWCNTs.

In Figure~\ref{fig:E11E22}(b), we directly compare the measured $E{22}$ transition energies from optical absorption spectroscopy\cite{OpticalAbsorbance} with our \LCAOTDDFTkomega{} calculated values including the derivative discontinuity correction of the \GLLBsc{} functional $\Deltax$.  Here we obtain a similar agreement to that for the $E_{11}$ transition, with an average error of $\epsilon \approx 26 \pm 33$~meV.  This is again well within the expected 0.1~eV accuracy of DFT calculations.  It is important to note that the $E_{11}$ and $E_{22}$ transition energies almost completely uncorrelated, as shown in Figure~\ref{fig:E11E22}, so that the near quantitative agreement we obtain is rather independent and systematic.

Based on the results of this subsection and the previous one, we have shown that the \LCAOTDDFTkomega{} code can reproduce with great accuracy the uncorrelated $E_{11}$ and $E_{22}$ transitions energies for SWCNTs. Moreover, the fact that we are obtaining such a good agreement suggests that we are considering almost all the processes that are taking place during the optical absorption. This also suggests that transitions that include charge transfer, which are not described by our method, do not occur in these SWCNTs.

\subsection{Electron Energy Loss Spectroscopy}\label{EELS}

So far we have considered the optical absorption spectra calculated using the \LCAOTDDFTkomega{} code, defined as the imaginary part of the dielectric function, $\Im[\varepsilon(\omega)]$, for semiconducting SWCNTs. Now we will consider the electron energy loss spectroscopy of both metallic and semiconducting SWCNTs, provided in Ref.~\citenum{SWCNTEELS}. In so doing we are able to also assess the accuracy of the real part of the dielectric function, $\Re[\varepsilon(\omega)]$. This is because the electron energy loss spectra is the negative of the imaginary part of the inverse of the dielectric function, $-\Im[\varepsilon^{-1}(\omega)]$, i.e., $\frac{\Im[\varepsilon(\omega)]}{{\Re[\varepsilon(\omega)]}^2+{\Im[\varepsilon(\omega)]}^2}$. In this way, we are further assessing the robustness of the \LCAOTDDFTkomega{} code by considering another of its outputs. Furthermore, the comparison will be done with respect to measured spectra that correspond to different experiments than those used in the previous sections.

\begin{figure}[!t]
  \includegraphics[width=\columnwidth]{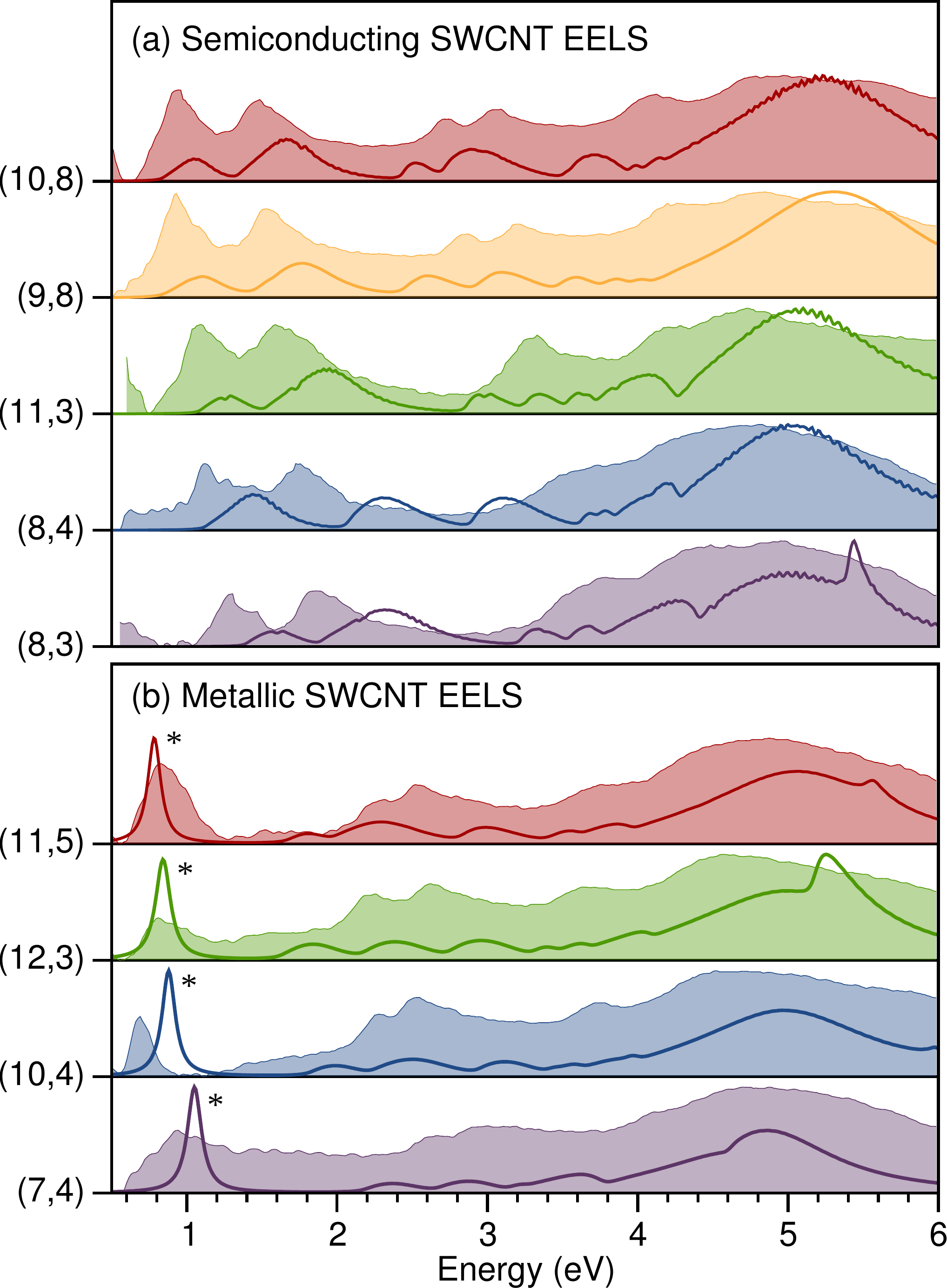}  
  \caption{Comparison of \LCAOTDDFTkomega{} calculated (solid lines) and measured (filled regions, Ref.~\citenum{SWCNTEELS}) electron energy loss $-\Im[\varepsilon^{-1}(\omega)]$ spectra along the SWCNT axis in eV for chirality sorted (a) semiconducting (10,8), (9,8), (11,3), (8,4), and (8,3) and (b) metallic (11,5), (12,3), (10,4) and (7,4) SWCNTs shown in Figure~\ref{fig:FigSWNTmap}, with Drude intraband plasmons $\omega_P$ (*) marked.}\label{fig:EELS}
\end{figure}

In Figure~\ref{fig:EELS}(a) we compare the electron energy loss spectra of semiconducting SWCNTs measured in Ref. \citenum{SWCNTEELS} with our calculations using the \LCAOTDDFTkomega{} code. We find for all five semiconducting SWCNTs that the first and second peaks are somewhat blue-shifted with respect to the measured spectra by about $0.2$ and $0.4$~eV on average, respectively. These peaks are assigned the $E_{11}$ and $E_{22}$ interband transitions.

Above these two peaks in energy there is a trough and one, two or three intermediate peaks before a broader and last peak in the measured spectra. The third of all the peaks, that is, the first of the intermediate peaks or the one right after the trough, is always red-shifted with respect to the measured spectra by about $0.37$~eV on average. These intermediate peaks correspond to the $E_{33}$, $E_{44}$ and $E_{55}$ interband transitions which can be easily identified in the spectra of the (10,8) and the (9,8) SWCNTs. 

The broader and higher energy peak is blue-shifted by about $0.29$~eV on average, and is the well-known $\pi$ plasmon of SWCNTs\cite{SWCNTpiEELS99,SWCNTpiEELS08,Silvina}. The spectra have the same behavior in general up to an energy shift, but in the spectra of the (8,4) and the (8,3) SWCNTs the peaks are closer together and harder to recognize. As in the experimental results, the spectra present a monotonic downshift as the diameter of the SWCNT increases.
 
Turning to an analysis of the metallic SWCNTs' electron energy loss spectroscopy in Figure~\ref{fig:EELS}(b), we observe a strong peak at around $1$~eV (marked with an *) that also matches what is observed in the experimental spectra. These peaks, which are present only in the spectra of metallic SWCNTs, correspond to free charge carrier Drude plasmons $\omega_P$. In other words, an intraband excitation that causes quantized collective oscillations of electrons. 

Going higher in energy, there is a trough and three well-known peaks. The first two of these peaks correspond to $M_{11}$ ($E_{11}$) transitions. The splitting of the transition into two peaks is probably caused by the trigonal wrapping effect \cite{SWCNTEELS}. The third peak corresponds to the $M_{22}$ transition. All of these transitions can be compared to peaks in the experimental data, although they are red-shifted by about $0.15$~eV. Finally, the broader and higher energy peak is again blue-shifted by about $0.28$~eV, and is the well-known $\pi$ plasmon of SWCNTs\cite{SWCNTpiEELS99,SWCNTpiEELS08,Silvina}. We also observe intense peaks in the spectra above $5$~eV of the (11,5) and (12,3) SWCNTs, which could be related to splitting of the $\pi$ plasmon.

In summary, we obtained an accurate energy for the plasmonic transition and also a qualitative description of the two peaks related to the $M_{11}$ transition and the peak related to the $M_{22}$ up to a red-shift. In this way we have assessed both the real and imaginary part of the dielectric function calculated using the \LCAOTDDFTkomega{} code and found it provides a robust and efficient method for modelling electron energy loss spectra.

\subsection{Exciton Density}\label{ExcitonDensity}

Having demonstrated the reliability of our \LCAOTDDFTkomega{} code for describing both the optical absorption and electron energy loss spectra of SWCNTs in the previous sections, we may now use the electron hole density difference $\Delta \rho(\rv,\omega)$, calculated from Eq.~\ref{eq:exciton_difference}, to model the spatial distribution of the most relevant bright excitons.  In so doing, we may probe the spatially resolved optical absorption and electron energy loss spectroscopy of SWCNTS, and their underlying physical makeup.

\begin{figure}
  \includegraphics[width=\columnwidth]{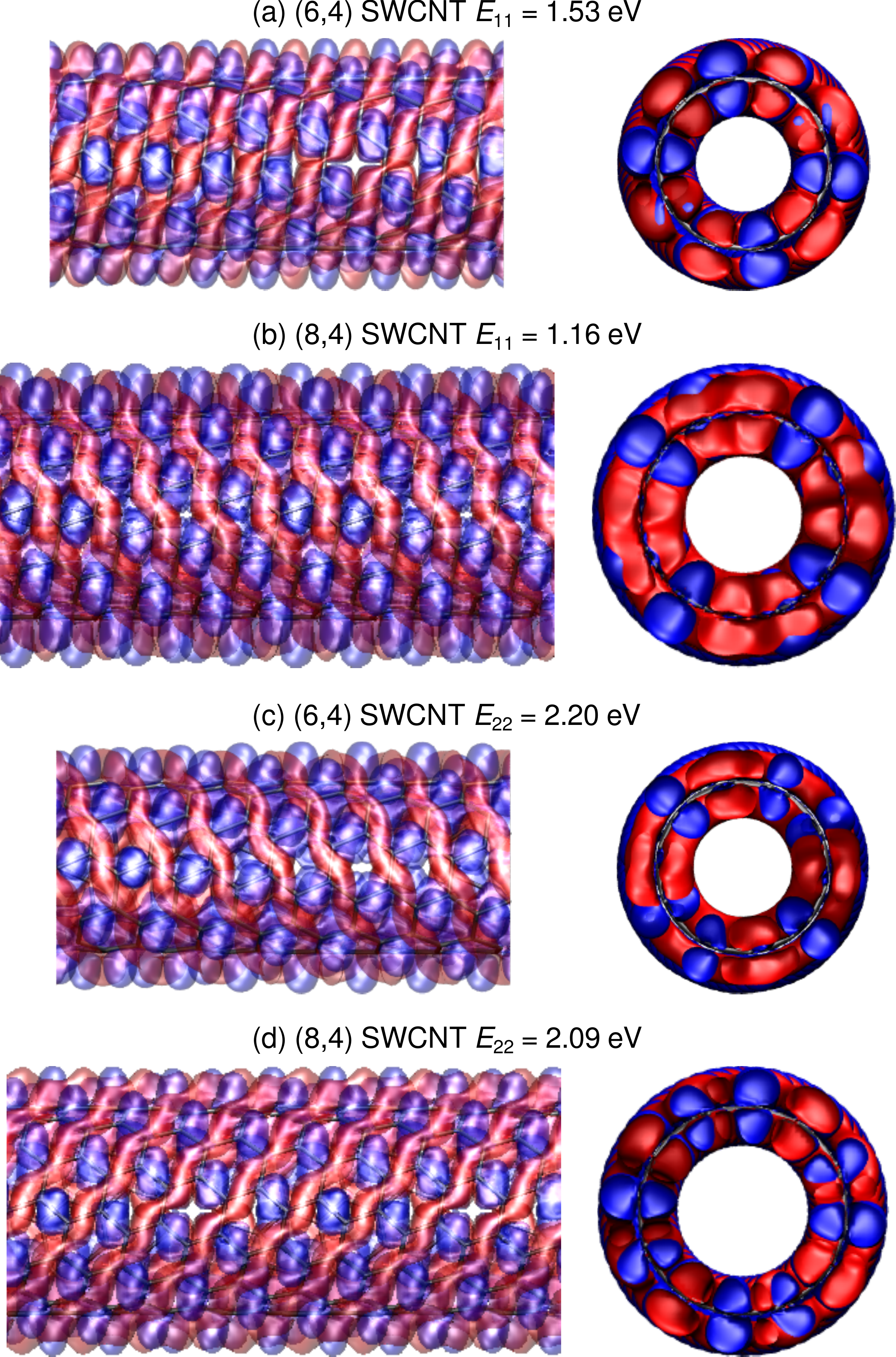}
  \caption{Exciton density difference $\Delta \rho(\rv, \omega) = \rho_e(\rv,\omega) + \rho_h(\rv,\omega)$ positive (red) and negative (blue) isosurfaces for the (a,b) $E_{11}$ and (c,d) $E_{22}$ transitions of the (a,c) (6,4) ($E_{11} \approx 1.53$, $E_{22}\approx2.20$~eV) and (b,d) (8,4) ($E_{11} \approx 1.16$, $E_{22} \approx 2.20$~eV) semiconducting SWCNTs along the axis (left) and in the plane (right) of the nanotube.}
  \label{fig:semiconductingexcitons}
\end{figure}

In Figure~\ref{fig:semiconductingexcitons} we show the spatially resolved electron-hole density difference $\Delta\rho(\rv,\omega)$ of the $E_{11}$ and $E_{22}$ transitions for two semiconducting SWCNTs with quite different transition energies.   Regions of negative charge (blue) correspond to the excited electron, whereas regions of positive charge (red) correspond to the hole. For both the $E_{11}$ and $E_{22}$ transitions we find that the positive (or hole) density is distributed in a continuous spiral around the nanotube, whereas the negative (or electron) density follows the same pattern but is discontinuous.  This suggests the hole density corresponds to bonding orbitals wrapping the SWCNT, whereas the electron density corresponds to anti-bonding orbitals localized on individual C--C bonds.

In fact, the plots in Figure~\ref{fig:semiconductingexcitons} in the SWCNT's plane show the electron hole density difference isosurfaces are composed of $\pi$-orbitals, with a nodal plane on the SWCNT's surface. These results clearly demonstrate that both the $E_{11}$ and $E_{22}$ transitions are indeed $\pi \to \pi$ transitions, as expected.  

It is interesting to note that the spatial distribution of the $E_{11}$ transition of the (6,4) SWCNT more closely resembles that of the $E_{22}$ transition of the (8,4) SWCNT, whereas the $E_{22}$ transition of the (6,4) SWCNT more closely resembles that of the $E_{11}$ transition of the (8,4) SWCNT.  This is evident from both the direction of the wrapping of the positive hole distributions around the nanotube axis and the slice in the nanotube plane.  This clearly suggests the spatial distribution of the individual excitonic peaks is highly dependent on the SWCNT's chirality, and not simply a function of the peak's energy.

\begin{figure}
\includegraphics[width=\columnwidth]{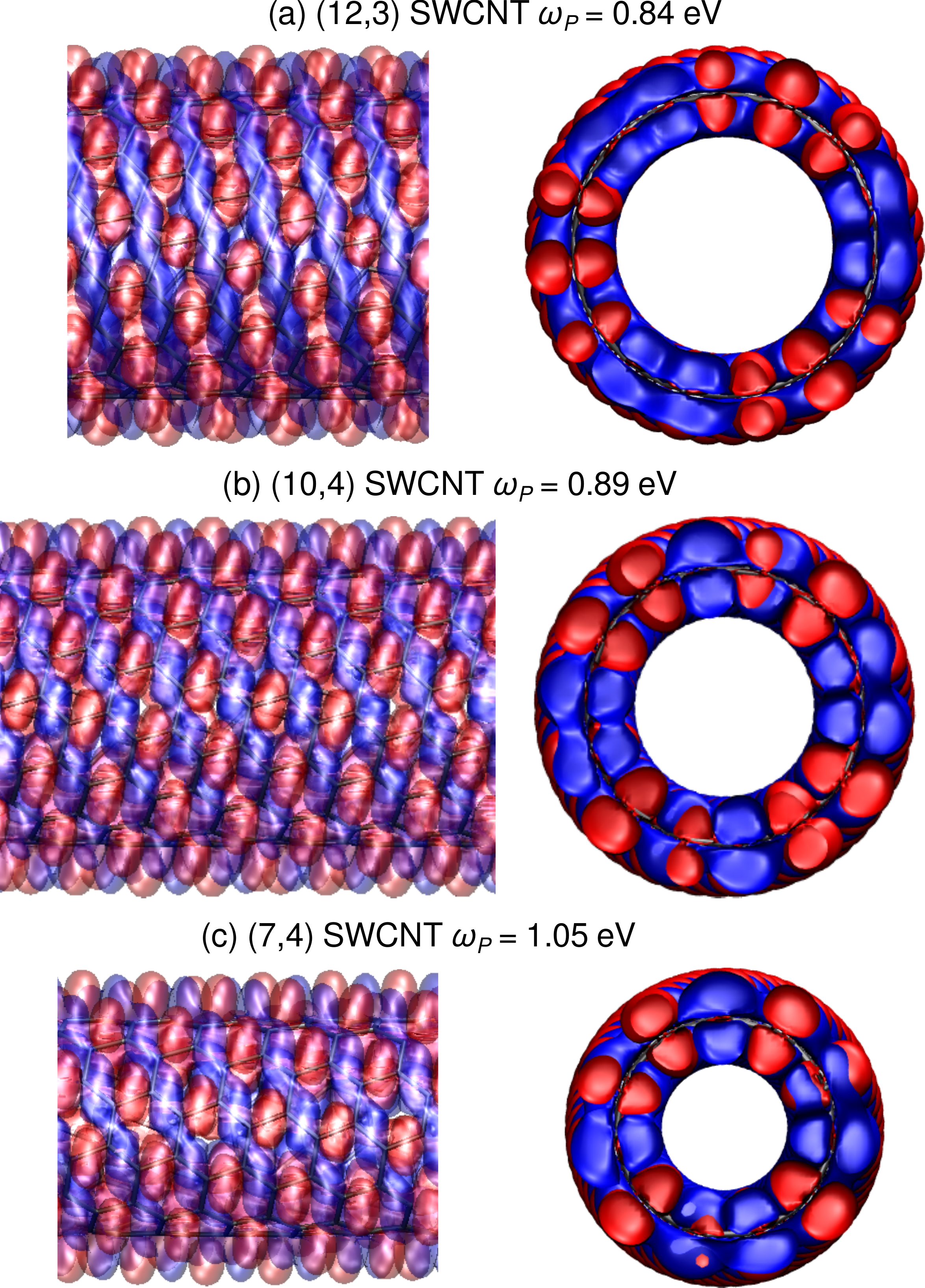}
  \caption{Exciton density difference $\Delta \rho(\rv, \omega) = \rho_e(\rv,\omega) + \rho_h(\rv,\omega)$ positive (red) and negative (blue) isosurfaces for the plasmon excitations $\omega_P$ of the (a) (12,3) ($\omega_P \approx 0.84$~eV), (b) (10,4) ($\omega_P \approx 0.89$~eV), and (c) (7,4) ($\omega_P \approx 1.05$~eV) metallic SWCNTs along the axis (left) and in the plane (right) of the nanotube.}
  \label{fig:metalexcitons}
\end{figure}

In Figure~\ref{fig:metalexcitons} we show the spatially resolved electron-hole density difference $\Delta\rho(\rv,\omega)$ of the Drude intraband plasmon $\omega_P$ for three different metallic SWCNTs.   Again, regions of negative charge (blue) correspond to the excited electron, whereas regions of positive charge (red) correspond to the hole. In contrast to the $E_{11}$ and $E_{22}$ transitions of the semiconducting SWCNTs (\emph{cf.}~Figure~\ref{fig:semiconductingexcitons}), we find for each of the three metallic nanotubes the plasmon excitation has negative (or electron) density distributed in a continuous spiral around the nanotube, whereas the positive (or hole) density follows the same pattern but is discontinuous.  This suggests the electron density corresponds to bonding orbitals wrapping the SWCNT, whereas the hole density corresponds to anti-bonding orbitals localized on individual C--C bonds.  

As was the case for the semiconducting $E_{11}$ and $E_{22}$ transitions (\emph{cf.}~Figure~\ref{fig:semiconductingexcitons}), the plots in Figure~\ref{fig:metalexcitons} in the SWCNT's plane show the electron hole density difference isosurfaces for the Drude intraband plasmons $\omega_P$ are also composed of $\pi$-orbitals, with a nodal plane on the SWCNT's surface. These results clearly demonstrate that the Drude plasmon is also composed of $\pi \to \pi$ transitions, as expected.  However, the nearly continuous excited electron's density seems to be a property of this metallic plasmon.

Overall, these results provide us with added insight into the physical makeup of the experimentally observed peaks in optical absorbance and electron energy loss spectra.  This information has the potential of further optimizing a SWCNT's overlap with donor molecules when designing organic photovoltaic cells.

\section{Conclusions}\label{Conclusions}

We have performed an in-depth analysis of the optical absorption and electron energy loss spectra of SWCNTs, 1D structures with properties determined by their $(m,n)$ chiral indices. We have considered a variety of SWCNTs with different indices and found that our theoretical optical absorption spectra, given by the imaginary part of the dielectric function, agree semi-quantitatively with the experimental data when the derivative discontinuity correction of the \GLLBsc{} functional $\Deltax$ is employed in our \LCAOTDDFTkomega{} code. We also see that both the calculated $E_{11}$ and $E_{22}$ transition energies have an average error much smaller than the expected accuracy of DFT calculations, with the $E_{22}$ transition energies having an even better agreement than the $E_{11}$. This result is rather surprising since the $E_{11}$ and $E_{22}$ energies are uncorrelated. Furthermore, we assessed the real part of the dielectric function by comparing our calculated electron energy loss spectra, given by minus the imaginary part of the inverse dielectric function, to experimental data. We were able to reproduce the qualitative behavior of the spectra and to obtain an accurate energy for the Drude intraband plasmon peak $\omega_P$ in metallic SWCNTs.  Finally, we have employed the electron hole density difference $\Delta \rho(\rv,\omega)$ to model the spatial distribution of the excitons.  We find, as expected, the $E_{11}$ and $E_{22}$ transitions in semiconducting SWCNTs and the Drude intraband plasmon $\omega_P$ in metallic SWCNTs all involve $\pi$ levels which wrap around the SWCNTs.    Altogether, these results demonstrate the surprising reliability and efficiency of a simplified LCAO-based TDDFT calculation in the optical limit for describing the optical absorbance and electron energy loss spectra of carbon-based macromolecules.  This work blazes the trail towards the computational design of complex carbon-based macromolecular organic photovoltaic systems \emph{in silico}.

\acknowledgments

This work employed the Imbabura cluster of Yachay Tech University, which was purchased under contract No.\ 2017-024 (SIE-UITEY-007-2017).


%\bibliography{thesis}

\begin{thebibliography}{49}%
\makeatletter
\providecommand \@ifxundefined [1]{%
 \@ifx{#1\undefined}
}%
\providecommand \@ifnum [1]{%
 \ifnum #1\expandafter \@firstoftwo
 \else \expandafter \@secondoftwo
 \fi
}%
\providecommand \@ifx [1]{%
 \ifx #1\expandafter \@firstoftwo
 \else \expandafter \@secondoftwo
 \fi
}%
\providecommand \natexlab [1]{#1}%
\providecommand \enquote  [1]{``#1''}%
\providecommand \bibnamefont  [1]{#1}%
\providecommand \bibfnamefont [1]{#1}%
\providecommand \citenamefont [1]{#1}%
\providecommand \href@noop [0]{\@secondoftwo}%
\providecommand \href [0]{\begingroup \@sanitize@url \@href}%
\providecommand \@href[1]{\@@startlink{#1}\@@href}%
\providecommand \@@href[1]{\endgroup#1\@@endlink}%
\providecommand \@sanitize@url [0]{\catcode `\\12\catcode `\$12\catcode
  `\&12\catcode `\#12\catcode `\^12\catcode `\_12\catcode `\%12\relax}%
\providecommand \@@startlink[1]{}%
\providecommand \@@endlink[0]{}%
\providecommand \url  [0]{\begingroup\@sanitize@url \@url }%
\providecommand \@url [1]{\endgroup\@href {#1}{\urlprefix }}%
\providecommand \urlprefix  [0]{URL }%
\providecommand \Eprint [0]{\href }%
\providecommand \doibase [0]{http://dx.doi.org/}%
\providecommand \selectlanguage [0]{\@gobble}%
\providecommand \bibinfo  [0]{\@secondoftwo}%
\providecommand \bibfield  [0]{\@secondoftwo}%
\providecommand \translation [1]{[#1]}%
\providecommand \BibitemOpen [0]{}%
\providecommand \bibitemStop [0]{}%
\providecommand \bibitemNoStop [0]{.\EOS\space}%
\providecommand \EOS [0]{\spacefactor3000\relax}%
\providecommand \BibitemShut  [1]{\csname bibitem#1\endcsname}%
\let\auto@bib@innerbib\@empty
%</preamble>
\bibitem [{\citenamefont {Baughman}\ \emph {et~al.}(2002)\citenamefont
  {Baughman}, \citenamefont {Zakhidov},\ and\ \citenamefont
  {de~Heer}}]{baughman2002carbon}%
  \BibitemOpen
  \bibfield  {author} {\bibinfo {author} {\bibfnamefont {R.~H.}\ \bibnamefont
  {Baughman}}, \bibinfo {author} {\bibfnamefont {A.~A.}\ \bibnamefont
  {Zakhidov}}, \ and\ \bibinfo {author} {\bibfnamefont {W.~A.}\ \bibnamefont
  {de~Heer}},\ }\bibfield  {title} {\enquote {\bibinfo {title} {Carbon
  nanotubes--the route toward applications},}\ }\href {\doibase
  10.1126/science.1060928} {\bibfield  {journal} {\bibinfo  {journal}
  {Science}\ }\textbf {\bibinfo {volume} {297}},\ \bibinfo {pages} {787--792}
  (\bibinfo {year} {2002})}\BibitemShut {NoStop}%
\bibitem [{\citenamefont {Dresselhaus}\ \emph {et~al.}(2001)\citenamefont
  {Dresselhaus}, \citenamefont {Dresselhaus},\ and\ \citenamefont
  {Avouris}}]{Dresselhaus}%
  \BibitemOpen
  \bibinfo {editor} {\bibfnamefont {M.~S.}\ \bibnamefont {Dresselhaus}},
  \bibinfo {editor} {\bibfnamefont {G.}~\bibnamefont {Dresselhaus}}, \ and\
  \bibinfo {editor} {\bibfnamefont {P.}~\bibnamefont {Avouris}},\ eds.,\
  \href@noop {} {\emph {\bibinfo {title} {Carbon Nanotubes: Synthesis,
  Structure, Properties, and Applications}}}\ (\bibinfo  {publisher}
  {Springer},\ \bibinfo {address} {Berlin},\ \bibinfo {year}
  {2001})\BibitemShut {NoStop}%
\bibitem [{\citenamefont {Spataru}\ \emph {et~al.}(2004)\citenamefont
  {Spataru}, \citenamefont {Ismail-Beigi}, \citenamefont {Benedict},\ and\
  \citenamefont {Louie}}]{SpataruPRL2004}%
  \BibitemOpen
  \bibfield  {author} {\bibinfo {author} {\bibfnamefont {C.~D.}\ \bibnamefont
  {Spataru}}, \bibinfo {author} {\bibfnamefont {S.}~\bibnamefont
  {Ismail-Beigi}}, \bibinfo {author} {\bibfnamefont {L.~X.}\ \bibnamefont
  {Benedict}}, \ and\ \bibinfo {author} {\bibfnamefont {S.~G.}\ \bibnamefont
  {Louie}},\ }\bibfield  {title} {\enquote {\bibinfo {title} {Excitonic effects
  and optical spectra of single-walled carbon nanotubes},}\ }\href {\doibase
  10.1103/PhysRevLett.92.077402} {\bibfield  {journal} {\bibinfo  {journal}
  {Phys. Rev. Lett.}\ }\textbf {\bibinfo {volume} {92}},\ \bibinfo {pages}
  {077402} (\bibinfo {year} {2004})}\BibitemShut {NoStop}%
\bibitem [{\citenamefont {Kataura}\ \emph {et~al.}(1999)\citenamefont
  {Kataura}, \citenamefont {Kumazawa}, \citenamefont {Maniwa}, \citenamefont
  {Umezu}, \citenamefont {Suzuki}, \citenamefont {Ohtsuka},\ and\ \citenamefont
  {Achiba}}]{kataura1999optical}%
  \BibitemOpen
  \bibfield  {author} {\bibinfo {author} {\bibfnamefont {H.}~\bibnamefont
  {Kataura}}, \bibinfo {author} {\bibfnamefont {Y.}~\bibnamefont {Kumazawa}},
  \bibinfo {author} {\bibfnamefont {Y.}~\bibnamefont {Maniwa}}, \bibinfo
  {author} {\bibfnamefont {I.}~\bibnamefont {Umezu}}, \bibinfo {author}
  {\bibfnamefont {S.}~\bibnamefont {Suzuki}}, \bibinfo {author} {\bibfnamefont
  {Y.}~\bibnamefont {Ohtsuka}}, \ and\ \bibinfo {author} {\bibfnamefont
  {Y.}~\bibnamefont {Achiba}},\ }\bibfield  {title} {\enquote {\bibinfo {title}
  {Optical properties of single-wall carbon nanotubes},}\ }\href {\doibase
  10.1016/S0379-6779(98)00278-1} {\bibfield  {journal} {\bibinfo  {journal}
  {Syn. Metals}\ }\textbf {\bibinfo {volume} {103}},\ \bibinfo {pages}
  {2555--2558} (\bibinfo {year} {1999})}\BibitemShut {NoStop}%
\bibitem [{\citenamefont {Yamamoto}\ \emph {et~al.}(2008)\citenamefont
  {Yamamoto}, \citenamefont {Watanabe},\ and\ \citenamefont
  {Hern{\'a}ndez}}]{Yamamoto2008}%
  \BibitemOpen
  \bibfield  {author} {\bibinfo {author} {\bibfnamefont {T.}~\bibnamefont
  {Yamamoto}}, \bibinfo {author} {\bibfnamefont {K.}~\bibnamefont {Watanabe}},
  \ and\ \bibinfo {author} {\bibfnamefont {E.~R.}\ \bibnamefont
  {Hern{\'a}ndez}},\ }\enquote {\bibinfo {title} {Mechanical properties,
  thermal stability and heat transport in carbon nanotubes},}\ in\ \href
  {\doibase 10.1007/978-3-540-72865-8_5} {\emph {\bibinfo {booktitle} {Carbon
  Nanotubes: Advanced Topics in the Synthesis, Structure, Properties and
  Applications}}},\ \bibinfo {editor} {edited by\ \bibinfo {editor}
  {\bibfnamefont {A.}~\bibnamefont {Jorio}}, \bibinfo {editor} {\bibfnamefont
  {G.}~\bibnamefont {Dresselhaus}}, \ and\ \bibinfo {editor} {\bibfnamefont
  {M.~S.}\ \bibnamefont {Dresselhaus}}}\ (\bibinfo  {publisher} {Springer
  Berlin Heidelberg},\ \bibinfo {address} {Berlin, Heidelberg},\ \bibinfo
  {year} {2008})\ pp.\ \bibinfo {pages} {165--195}\BibitemShut {NoStop}%
\bibitem [{\citenamefont {Liew}\ \emph {et~al.}(2005)\citenamefont {Liew},
  \citenamefont {Wong}, \citenamefont {He},\ and\ \citenamefont
  {Tan}}]{Liew2005Thermal}%
  \BibitemOpen
  \bibfield  {author} {\bibinfo {author} {\bibfnamefont {K.~M.}\ \bibnamefont
  {Liew}}, \bibinfo {author} {\bibfnamefont {C.~H.}\ \bibnamefont {Wong}},
  \bibinfo {author} {\bibfnamefont {X.~Q.}\ \bibnamefont {He}}, \ and\ \bibinfo
  {author} {\bibfnamefont {M.~J.}\ \bibnamefont {Tan}},\ }\bibfield  {title}
  {\enquote {\bibinfo {title} {Thermal stability of single and multi-walled
  carbon nanotubes},}\ }\href {\doibase 10.1103/PhysRevB.71.075424} {\bibfield
  {journal} {\bibinfo  {journal} {Phys. Rev. B}\ }\textbf {\bibinfo {volume}
  {71}},\ \bibinfo {pages} {075424} (\bibinfo {year} {2005})}\BibitemShut
  {NoStop}%
\bibitem [{\citenamefont {Liang}\ \emph {et~al.}(2001)\citenamefont {Liang},
  \citenamefont {Bockrath}, \citenamefont {Bozovic}, \citenamefont {Hafner},
  \citenamefont {Tinkham},\ and\ \citenamefont {Park}}]{liang2001fabry}%
  \BibitemOpen
  \bibfield  {author} {\bibinfo {author} {\bibfnamefont {W.}~\bibnamefont
  {Liang}}, \bibinfo {author} {\bibfnamefont {M.}~\bibnamefont {Bockrath}},
  \bibinfo {author} {\bibfnamefont {D.}~\bibnamefont {Bozovic}}, \bibinfo
  {author} {\bibfnamefont {J.~H.}\ \bibnamefont {Hafner}}, \bibinfo {author}
  {\bibfnamefont {M.}~\bibnamefont {Tinkham}}, \ and\ \bibinfo {author}
  {\bibfnamefont {H.}~\bibnamefont {Park}},\ }\bibfield  {title} {\enquote
  {\bibinfo {title} {{Fabry-Perot} interference in a nanotube electron
  waveguide},}\ }\href {\doibase 10.1038/35079517} {\bibfield  {journal}
  {\bibinfo  {journal} {Nature}\ }\textbf {\bibinfo {volume} {411}},\ \bibinfo
  {pages} {665--669} (\bibinfo {year} {2001})}\BibitemShut {NoStop}%
\bibitem [{\citenamefont {Frank}\ \emph {et~al.}(1998)\citenamefont {Frank},
  \citenamefont {Poncharal}, \citenamefont {Wang},\ and\ \citenamefont
  {De~Heer}}]{frank1998carbon}%
  \BibitemOpen
  \bibfield  {author} {\bibinfo {author} {\bibfnamefont {S.}~\bibnamefont
  {Frank}}, \bibinfo {author} {\bibfnamefont {P.}~\bibnamefont {Poncharal}},
  \bibinfo {author} {\bibfnamefont {Z.}~\bibnamefont {Wang}}, \ and\ \bibinfo
  {author} {\bibfnamefont {W.~A.}\ \bibnamefont {De~Heer}},\ }\bibfield
  {title} {\enquote {\bibinfo {title} {Carbon nanotube quantum resistors},}\
  }\href {\doibase 10.1126/science.280.5370.1744} {\bibfield  {journal}
  {\bibinfo  {journal} {Science}\ }\textbf {\bibinfo {volume} {280}},\ \bibinfo
  {pages} {1744--1746} (\bibinfo {year} {1998})}\BibitemShut {NoStop}%
\bibitem [{\citenamefont {Kymakis}\ and\ \citenamefont
  {Amaratunga}(2002)}]{kymakis2002single}%
  \BibitemOpen
  \bibfield  {author} {\bibinfo {author} {\bibfnamefont {E.}~\bibnamefont
  {Kymakis}}\ and\ \bibinfo {author} {\bibfnamefont {G.}~\bibnamefont
  {Amaratunga}},\ }\bibfield  {title} {\enquote {\bibinfo {title} {Single-wall
  carbon nanotube/conjugated polymer photovoltaic devices},}\ }\href {\doibase
  10.1063/1.1428416} {\bibfield  {journal} {\bibinfo  {journal} {Appl. Phys.
  Lett.}\ }\textbf {\bibinfo {volume} {80}},\ \bibinfo {pages} {112--114}
  (\bibinfo {year} {2002})}\BibitemShut {NoStop}%
\bibitem [{\citenamefont {Kymakis}\ \emph {et~al.}(2006)\citenamefont
  {Kymakis}, \citenamefont {Koudoumas}, \citenamefont {Franghiadakis},\ and\
  \citenamefont {Amaratunga}}]{kymakis2006post}%
  \BibitemOpen
  \bibfield  {author} {\bibinfo {author} {\bibfnamefont {E.}~\bibnamefont
  {Kymakis}}, \bibinfo {author} {\bibfnamefont {E.}~\bibnamefont {Koudoumas}},
  \bibinfo {author} {\bibfnamefont {I.}~\bibnamefont {Franghiadakis}}, \ and\
  \bibinfo {author} {\bibfnamefont {G.}~\bibnamefont {Amaratunga}},\ }\bibfield
   {title} {\enquote {\bibinfo {title} {Post-fabrication annealing effects in
  polymer-nanotube photovoltaic cells},}\ }\href {\doibase
  10.1088/0022-3727/39/6/010} {\bibfield  {journal} {\bibinfo  {journal} {J.
  Phys. D Appl. Phys.}\ }\textbf {\bibinfo {volume} {39}},\ \bibinfo {pages}
  {1058} (\bibinfo {year} {2006})}\BibitemShut {NoStop}%
\bibitem [{\citenamefont {Campidelli}\ \emph {et~al.}(2008)\citenamefont
  {Campidelli}, \citenamefont {Ballesteros}, \citenamefont {Filoramo},
  \citenamefont {D{\'i}az}, \citenamefont {de~la Torre}, \citenamefont
  {Torres}, \citenamefont {Rahman}, \citenamefont {Ehli}, \citenamefont
  {Kiessling},\ and\ \citenamefont {Werner}}]{campidelli2008facile}%
  \BibitemOpen
  \bibfield  {author} {\bibinfo {author} {\bibfnamefont {S.}~\bibnamefont
  {Campidelli}}, \bibinfo {author} {\bibfnamefont {B.}~\bibnamefont
  {Ballesteros}}, \bibinfo {author} {\bibfnamefont {A.}~\bibnamefont
  {Filoramo}}, \bibinfo {author} {\bibfnamefont {D.}~\bibnamefont {D{\'i}az}},
  \bibinfo {author} {\bibfnamefont {G.}~\bibnamefont {de~la Torre}}, \bibinfo
  {author} {\bibfnamefont {T.}~\bibnamefont {Torres}}, \bibinfo {author}
  {\bibfnamefont {G.~A.}\ \bibnamefont {Rahman}}, \bibinfo {author}
  {\bibfnamefont {C.}~\bibnamefont {Ehli}}, \bibinfo {author} {\bibfnamefont
  {D.}~\bibnamefont {Kiessling}}, \ and\ \bibinfo {author} {\bibfnamefont
  {F.}~\bibnamefont {Werner}},\ }\bibfield  {title} {\enquote {\bibinfo {title}
  {Facile decoration of functionalized single-wall carbon nanotubes with
  phthalocyanines via ``click chemistry''},}\ }\href {\doibase
  10.1021/ja8033262} {\bibfield  {journal} {\bibinfo  {journal} {J. Am. Chem.
  Soc.}\ }\textbf {\bibinfo {volume} {130}},\ \bibinfo {pages} {11503--11509}
  (\bibinfo {year} {2008})}\BibitemShut {NoStop}%
\bibitem [{\citenamefont {Bartelmess}\ \emph {et~al.}(2010)\citenamefont
  {Bartelmess}, \citenamefont {Ballesteros}, \citenamefont {de~la Torre},
  \citenamefont {Kiessling}, \citenamefont {Campidelli}, \citenamefont {Prato},
  \citenamefont {Torres},\ and\ \citenamefont
  {Guldi}}]{bartelmess2010phthalocyanine}%
  \BibitemOpen
  \bibfield  {author} {\bibinfo {author} {\bibfnamefont {J.}~\bibnamefont
  {Bartelmess}}, \bibinfo {author} {\bibfnamefont {B.}~\bibnamefont
  {Ballesteros}}, \bibinfo {author} {\bibfnamefont {G.}~\bibnamefont {de~la
  Torre}}, \bibinfo {author} {\bibfnamefont {D.}~\bibnamefont {Kiessling}},
  \bibinfo {author} {\bibfnamefont {S.}~\bibnamefont {Campidelli}}, \bibinfo
  {author} {\bibfnamefont {M.}~\bibnamefont {Prato}}, \bibinfo {author}
  {\bibfnamefont {T.}~\bibnamefont {Torres}}, \ and\ \bibinfo {author}
  {\bibfnamefont {D.~M.}\ \bibnamefont {Guldi}},\ }\bibfield  {title} {\enquote
  {\bibinfo {title} {Phthalocyanine-pyrene conjugates: a powerful approach
  toward carbon nanotube solar cells},}\ }\href {\doibase 10.1021/ja107131r}
  {\bibfield  {journal} {\bibinfo  {journal} {J. Am. Chem. Soc.}\ }\textbf
  {\bibinfo {volume} {132}},\ \bibinfo {pages} {16202--16211} (\bibinfo {year}
  {2010})}\BibitemShut {NoStop}%
\bibitem [{\citenamefont {Guldi}\ \emph {et~al.}(2005)\citenamefont {Guldi},
  \citenamefont {Rahman}, \citenamefont {Prato}, \citenamefont {Jux},
  \citenamefont {Qin},\ and\ \citenamefont {Ford}}]{guldi2005single}%
  \BibitemOpen
  \bibfield  {author} {\bibinfo {author} {\bibfnamefont {D.~M.}\ \bibnamefont
  {Guldi}}, \bibinfo {author} {\bibfnamefont {G.}~\bibnamefont {Rahman}},
  \bibinfo {author} {\bibfnamefont {M.}~\bibnamefont {Prato}}, \bibinfo
  {author} {\bibfnamefont {N.}~\bibnamefont {Jux}}, \bibinfo {author}
  {\bibfnamefont {S.}~\bibnamefont {Qin}}, \ and\ \bibinfo {author}
  {\bibfnamefont {W.}~\bibnamefont {Ford}},\ }\bibfield  {title} {\enquote
  {\bibinfo {title} {Single-wall carbon nanotubes as integrative building
  blocks for solar-energy conversion},}\ }\href {\doibase
  10.1002/anie.200462416} {\bibfield  {journal} {\bibinfo  {journal} {Angew.
  Chem. Int. Ed.}\ }\textbf {\bibinfo {volume} {44}},\ \bibinfo {pages}
  {2015--2018} (\bibinfo {year} {2005})}\BibitemShut {NoStop}%
\bibitem [{\citenamefont {Ham}\ \emph {et~al.}(2010)\citenamefont {Ham},
  \citenamefont {Choi}, \citenamefont {Boghossian}, \citenamefont {Jeng},
  \citenamefont {Graff}, \citenamefont {Heller}, \citenamefont {Chang},
  \citenamefont {Mattis}, \citenamefont {Bayburt}, \citenamefont {Grinkova},
  \citenamefont {Zeiger}, \citenamefont {Vliet}, \citenamefont {Hobbie},
  \citenamefont {Sligar}, \citenamefont {Wraight},\ and\ \citenamefont
  {Strano}}]{Ham2010Photoelectrochemical}%
  \BibitemOpen
  \bibfield  {author} {\bibinfo {author} {\bibfnamefont {M.-H.}\ \bibnamefont
  {Ham}}, \bibinfo {author} {\bibfnamefont {J.~H.}\ \bibnamefont {Choi}},
  \bibinfo {author} {\bibfnamefont {A.~A.}\ \bibnamefont {Boghossian}},
  \bibinfo {author} {\bibfnamefont {E.~S.}\ \bibnamefont {Jeng}}, \bibinfo
  {author} {\bibfnamefont {R.~A.}\ \bibnamefont {Graff}}, \bibinfo {author}
  {\bibfnamefont {D.~A.}\ \bibnamefont {Heller}}, \bibinfo {author}
  {\bibfnamefont {A.~C.}\ \bibnamefont {Chang}}, \bibinfo {author}
  {\bibfnamefont {A.}~\bibnamefont {Mattis}}, \bibinfo {author} {\bibfnamefont
  {T.~H.}\ \bibnamefont {Bayburt}}, \bibinfo {author} {\bibfnamefont {Y.~V.}\
  \bibnamefont {Grinkova}}, \bibinfo {author} {\bibfnamefont {A.~S.}\
  \bibnamefont {Zeiger}}, \bibinfo {author} {\bibfnamefont {K.~J.~V.}\
  \bibnamefont {Vliet}}, \bibinfo {author} {\bibfnamefont {E.~K.}\ \bibnamefont
  {Hobbie}}, \bibinfo {author} {\bibfnamefont {S.~G.}\ \bibnamefont {Sligar}},
  \bibinfo {author} {\bibfnamefont {C.~A.}\ \bibnamefont {Wraight}}, \ and\
  \bibinfo {author} {\bibfnamefont {M.~S.}\ \bibnamefont {Strano}},\ }\bibfield
   {title} {\enquote {\bibinfo {title} {Photoelectrochemical complexes for
  solar energy conversion that chemically and autonomously regenerate},}\
  }\href {\doibase 10.1038/nchem.822} {\bibfield  {journal} {\bibinfo
  {journal} {Nat. Chem.}\ }\textbf {\bibinfo {volume} {2}},\ \bibinfo {pages}
  {929--936} (\bibinfo {year} {2010})}\BibitemShut {NoStop}%
\bibitem [{\citenamefont {Zamora-Ledezma}\ \emph {et~al.}(2009)\citenamefont
  {Zamora-Ledezma}, \citenamefont {Blanc},\ and\ \citenamefont
  {Anglaret}}]{ZamoraLedezma2009Orientational}%
  \BibitemOpen
  \bibfield  {author} {\bibinfo {author} {\bibfnamefont {C.}~\bibnamefont
  {Zamora-Ledezma}}, \bibinfo {author} {\bibfnamefont {C.}~\bibnamefont
  {Blanc}}, \ and\ \bibinfo {author} {\bibfnamefont {E.}~\bibnamefont
  {Anglaret}},\ }\bibfield  {title} {\enquote {\bibinfo {title} {Orientational
  order of single-wall carbon nanotubes in stretch-aligned photoluminescent
  composite films},}\ }\href {\doibase 10.1103/physrevb.80.113407} {\bibfield
  {journal} {\bibinfo  {journal} {Phys. Rev. B}\ }\textbf {\bibinfo {volume}
  {80}},\ \bibinfo {pages} {113407} (\bibinfo {year} {2009})}\BibitemShut
  {NoStop}%
\bibitem [{\citenamefont {Torres-Canas}\ \emph {et~al.}(2014)\citenamefont
  {Torres-Canas}, \citenamefont {Blanc}, \citenamefont {Zamora-Ledezma},
  \citenamefont {Silva},\ and\ \citenamefont
  {Anglaret}}]{TorresCanas2014Dispersion}%
  \BibitemOpen
  \bibfield  {author} {\bibinfo {author} {\bibfnamefont {F.~J.}\ \bibnamefont
  {Torres-Canas}}, \bibinfo {author} {\bibfnamefont {C.}~\bibnamefont {Blanc}},
  \bibinfo {author} {\bibfnamefont {C.}~\bibnamefont {Zamora-Ledezma}},
  \bibinfo {author} {\bibfnamefont {P.}~\bibnamefont {Silva}}, \ and\ \bibinfo
  {author} {\bibfnamefont {E.}~\bibnamefont {Anglaret}},\ }\bibfield  {title}
  {\enquote {\bibinfo {title} {Dispersion and individualization of {SWNT} in
  surfactant-free suspensions and composites of hydrosoluble polymers},}\
  }\href {\doibase 10.1021/jp5092015} {\bibfield  {journal} {\bibinfo
  {journal} {J. Phys. Chem. C}\ }\textbf {\bibinfo {volume} {119}},\ \bibinfo
  {pages} {703--709} (\bibinfo {year} {2014})}\BibitemShut {NoStop}%
\bibitem [{\citenamefont {Weisman}\ and\ \citenamefont
  {Kono}(2019)}]{Weisman2019Introduction}%
  \BibitemOpen
  \bibfield  {author} {\bibinfo {author} {\bibfnamefont {R.~B.}\ \bibnamefont
  {Weisman}}\ and\ \bibinfo {author} {\bibfnamefont {J.}~\bibnamefont {Kono}},\
  }\bibfield  {title} {\enquote {\bibinfo {title} {Introduction to optical
  spectroscopy of single-wall carbon nanotubes},}\ }in\ \href {\doibase
  10.1142/9789813235465_0001} {\emph {\bibinfo {booktitle} {Handbook of Carbon
  Nanomaterials}}}\ (\bibinfo  {publisher} {World Scientific},\ \bibinfo {year}
  {2019})\ pp.\ \bibinfo {pages} {1--43}\BibitemShut {NoStop}%
\bibitem [{\citenamefont {Zangwill}(2015)}]{Zangwill2015A}%
  \BibitemOpen
  \bibfield  {author} {\bibinfo {author} {\bibfnamefont {A.}~\bibnamefont
  {Zangwill}},\ }\bibfield  {title} {\enquote {\bibinfo {title} {A half century
  of density functional theory},}\ }\href {\doibase 10.1063/pt.3.2846}
  {\bibfield  {journal} {\bibinfo  {journal} {Physics Today}\ }\textbf
  {\bibinfo {volume} {68}},\ \bibinfo {pages} {34--39} (\bibinfo {year}
  {2015})}\BibitemShut {NoStop}%
\bibitem [{\citenamefont {N{\o}rskov}\ \emph {et~al.}(2009)\citenamefont
  {N{\o}rskov}, \citenamefont {Bligaard}, \citenamefont {Rossmeisl},\ and\
  \citenamefont {Christensen}}]{Nrskov2009Towards}%
  \BibitemOpen
  \bibfield  {author} {\bibinfo {author} {\bibfnamefont {J.~K.}\ \bibnamefont
  {N{\o}rskov}}, \bibinfo {author} {\bibfnamefont {T.}~\bibnamefont
  {Bligaard}}, \bibinfo {author} {\bibfnamefont {J.}~\bibnamefont {Rossmeisl}},
  \ and\ \bibinfo {author} {\bibfnamefont {C.~H.}\ \bibnamefont
  {Christensen}},\ }\bibfield  {title} {\enquote {\bibinfo {title} {Towards the
  computational design of solid catalysts},}\ }\href {\doibase
  10.1038/nchem.121} {\bibfield  {journal} {\bibinfo  {journal} {Nat. Chem.}\
  }\textbf {\bibinfo {volume} {1}},\ \bibinfo {pages} {37--46} (\bibinfo {year}
  {2009})}\BibitemShut {NoStop}%
\bibitem [{\citenamefont {Jain}\ \emph {et~al.}(2013)\citenamefont {Jain},
  \citenamefont {Ong}, \citenamefont {Hautier}, \citenamefont {Chen},
  \citenamefont {Richards}, \citenamefont {Dacek}, \citenamefont {Cholia},
  \citenamefont {Gunter}, \citenamefont {Skinner}, \citenamefont {Ceder},\ and\
  \citenamefont {Persson}}]{Jain2013Commentary}%
  \BibitemOpen
  \bibfield  {author} {\bibinfo {author} {\bibfnamefont {A.}~\bibnamefont
  {Jain}}, \bibinfo {author} {\bibfnamefont {S.~P.}\ \bibnamefont {Ong}},
  \bibinfo {author} {\bibfnamefont {G.}~\bibnamefont {Hautier}}, \bibinfo
  {author} {\bibfnamefont {W.}~\bibnamefont {Chen}}, \bibinfo {author}
  {\bibfnamefont {W.~D.}\ \bibnamefont {Richards}}, \bibinfo {author}
  {\bibfnamefont {S.}~\bibnamefont {Dacek}}, \bibinfo {author} {\bibfnamefont
  {S.}~\bibnamefont {Cholia}}, \bibinfo {author} {\bibfnamefont
  {D.}~\bibnamefont {Gunter}}, \bibinfo {author} {\bibfnamefont
  {D.}~\bibnamefont {Skinner}}, \bibinfo {author} {\bibfnamefont
  {G.}~\bibnamefont {Ceder}}, \ and\ \bibinfo {author} {\bibfnamefont {K.~A.}\
  \bibnamefont {Persson}},\ }\bibfield  {title} {\enquote {\bibinfo {title}
  {Commentary: The materials project: A materials genome approach to
  accelerating materials innovation},}\ }\href {\doibase 10.1063/1.4812323}
  {\bibfield  {journal} {\bibinfo  {journal} {{APL} Materials}\ }\textbf
  {\bibinfo {volume} {1}},\ \bibinfo {pages} {011002} (\bibinfo {year}
  {2013})}\BibitemShut {NoStop}%
\bibitem [{\citenamefont {Glanzmann}\ \emph {et~al.}(2015)\citenamefont
  {Glanzmann}, \citenamefont {Mowbray}, \citenamefont {{Figueroa del Valle}},
  \citenamefont {Scotognella}, \citenamefont {Lanzani},\ and\ \citenamefont
  {Rubio}}]{LiviaMilan}%
  \BibitemOpen
  \bibfield  {author} {\bibinfo {author} {\bibfnamefont {L.~N.}\ \bibnamefont
  {Glanzmann}}, \bibinfo {author} {\bibfnamefont {D.~J.}\ \bibnamefont
  {Mowbray}}, \bibinfo {author} {\bibfnamefont {D.~G.}\ \bibnamefont {{Figueroa
  del Valle}}}, \bibinfo {author} {\bibfnamefont {F.}~\bibnamefont
  {Scotognella}}, \bibinfo {author} {\bibfnamefont {G.}~\bibnamefont
  {Lanzani}}, \ and\ \bibinfo {author} {\bibfnamefont {A.}~\bibnamefont
  {Rubio}},\ }\bibfield  {title} {\enquote {\bibinfo {title} {Photoinduced
  absorption within single-walled carbon nanotube systems},}\ }\href {\doibase
  10.1021/acs.jpcc.5b10025} {\bibfield  {journal} {\bibinfo  {journal} {J.
  Phys. Chem. C}\ }\textbf {\bibinfo {volume} {120}},\ \bibinfo {pages}
  {1926--1935} (\bibinfo {year} {2015})}\BibitemShut {NoStop}%
\bibitem [{\citenamefont {Glanzmann}\ and\ \citenamefont
  {Mowbray}(2016)}]{Glanzmann2016Theoretical}%
  \BibitemOpen
  \bibfield  {author} {\bibinfo {author} {\bibfnamefont {L.~N.}\ \bibnamefont
  {Glanzmann}}\ and\ \bibinfo {author} {\bibfnamefont {D.~J.}\ \bibnamefont
  {Mowbray}},\ }\bibfield  {title} {\enquote {\bibinfo {title} {Theoretical
  insight into the internal quantum efficiencies of polymer/{C60} and
  polymer/{SWNT} photovoltaic devices},}\ }\href {\doibase
  10.1021/acs.jpcc.5b12611} {\bibfield  {journal} {\bibinfo  {journal} {J.
  Phys. Chem. C}\ }\textbf {\bibinfo {volume} {120}},\ \bibinfo {pages}
  {6336--6343} (\bibinfo {year} {2016})}\BibitemShut {NoStop}%
\bibitem [{\citenamefont {Wei}\ \emph {et~al.}(2016)\citenamefont {Wei},
  \citenamefont {Tanaka}, \citenamefont {Yomogida}, \citenamefont {Sato},
  \citenamefont {Saito},\ and\ \citenamefont {Kataura}}]{OpticalAbsorbance}%
  \BibitemOpen
  \bibfield  {author} {\bibinfo {author} {\bibfnamefont {X.}~\bibnamefont
  {Wei}}, \bibinfo {author} {\bibfnamefont {T.}~\bibnamefont {Tanaka}},
  \bibinfo {author} {\bibfnamefont {Y.}~\bibnamefont {Yomogida}}, \bibinfo
  {author} {\bibfnamefont {N.}~\bibnamefont {Sato}}, \bibinfo {author}
  {\bibfnamefont {R.}~\bibnamefont {Saito}}, \ and\ \bibinfo {author}
  {\bibfnamefont {H.}~\bibnamefont {Kataura}},\ }\bibfield  {title} {\enquote
  {\bibinfo {title} {Experimental determination of excitonic band structures of
  single-walled carbon nanotubes using circular dichroism spectra},}\ }\href
  {\doibase 10.1038/ncomms12899} {\bibfield  {journal} {\bibinfo  {journal}
  {Nat. Comm.}\ }\textbf {\bibinfo {volume} {7}},\ \bibinfo {pages} {12899}
  (\bibinfo {year} {2016})}\BibitemShut {NoStop}%
\bibitem [{\citenamefont {Senga}\ \emph {et~al.}(2016)\citenamefont {Senga},
  \citenamefont {Pichler},\ and\ \citenamefont {Suenaga}}]{SWCNTEELS}%
  \BibitemOpen
  \bibfield  {author} {\bibinfo {author} {\bibfnamefont {R.}~\bibnamefont
  {Senga}}, \bibinfo {author} {\bibfnamefont {T.}~\bibnamefont {Pichler}}, \
  and\ \bibinfo {author} {\bibfnamefont {K.}~\bibnamefont {Suenaga}},\
  }\bibfield  {title} {\enquote {\bibinfo {title} {Electron spectroscopy of
  single quantum objects to directly correlate the local structure to their
  electronic transport and optical properties},}\ }\href {\doibase
  10.1021/acs.nanolett.6b00825} {\bibfield  {journal} {\bibinfo  {journal}
  {Nano Lett.}\ }\textbf {\bibinfo {volume} {16}},\ \bibinfo {pages}
  {3661--3667} (\bibinfo {year} {2016})}\BibitemShut {NoStop}%
\bibitem [{\citenamefont {Weisman}\ and\ \citenamefont
  {Subramoney}(2006)}]{Weisman2006vanHoveE}%
  \BibitemOpen
  \bibfield  {author} {\bibinfo {author} {\bibfnamefont {R.}~\bibnamefont
  {Weisman}}\ and\ \bibinfo {author} {\bibfnamefont {S.}~\bibnamefont
  {Subramoney}},\ }\bibfield  {title} {\enquote {\bibinfo {title} {Carbon
  nanotubes},}\ }\href
  {https://www.electrochem.org/dl/interface/sum/sum06/sum06_p42.pdf} {\bibfield
   {journal} {\bibinfo  {journal} {Electrochem. Soc. Interface}\ }\textbf
  {\bibinfo {volume} {15}},\ \bibinfo {pages} {42--46} (\bibinfo {year}
  {2006})}\BibitemShut {NoStop}%
\bibitem [{\citenamefont {Lyon}\ \emph {et~al.}(2019)\citenamefont {Lyon},
  \citenamefont {Preciado-Rivas}, \citenamefont {Despoja},\ and\ \citenamefont
  {Mowbray}}]{LCAOTDDFTKeenan}%
  \BibitemOpen
  \bibfield  {author} {\bibinfo {author} {\bibfnamefont {K.}~\bibnamefont
  {Lyon}}, \bibinfo {author} {\bibfnamefont {M.~R.}\ \bibnamefont
  {Preciado-Rivas}}, \bibinfo {author} {\bibfnamefont {V.}~\bibnamefont
  {Despoja}}, \ and\ \bibinfo {author} {\bibfnamefont {D.~J.}\ \bibnamefont
  {Mowbray}},\ }\bibfield  {title} {\enquote {\bibinfo {title}
  {{\LCAOTDDFTkomega{}}: Spectroscopy in the optical limit},}\ }\href@noop {}
  {\  (\bibinfo {year} {2019})},\ \bibinfo {note} {unpublished}\BibitemShut
  {NoStop}%
\bibitem [{\citenamefont {Preciado-Rivas}\ \emph {et~al.}(2019)\citenamefont
  {Preciado-Rivas}, \citenamefont {Mowbray}, \citenamefont {Lyon},
  \citenamefont {Larsen},\ and\ \citenamefont {Milne}}]{Chlorophyll}%
  \BibitemOpen
  \bibfield  {author} {\bibinfo {author} {\bibfnamefont {M.~R.}\ \bibnamefont
  {Preciado-Rivas}}, \bibinfo {author} {\bibfnamefont {D.~J.}\ \bibnamefont
  {Mowbray}}, \bibinfo {author} {\bibfnamefont {K.}~\bibnamefont {Lyon}},
  \bibinfo {author} {\bibfnamefont {A.~H.}\ \bibnamefont {Larsen}}, \ and\
  \bibinfo {author} {\bibfnamefont {B.~F.}\ \bibnamefont {Milne}},\ }\bibfield
  {title} {\enquote {\bibinfo {title} {Optical excitations of chlorophyll
  \emph{a} and \emph{b} monomers and dimers},}\ }\href@noop {} {\  (\bibinfo
  {year} {2019})},\ \bibinfo {note}
  {arXiv:\href{arXiv.org/1907.09430}{1907.09430}}\BibitemShut {NoStop}%
\bibitem [{LCA()}]{LCAOTDDFTkomega}%
  \BibitemOpen
  \href@noop {} {}\bibinfo {note} {{T}he \LCAOTDDFTkomega{} code is available
  free of charge from
  \href{https://gitlab.com/lcao-tddft-k-omega/lcao-tddft-k-omega}{gitlab.com/lcao-tddft-k-omega/lcao-tddft-k-omega}}\BibitemShut
  {NoStop}%
\bibitem [{\citenamefont {Kuisma}\ \emph {et~al.}(2010)\citenamefont {Kuisma},
  \citenamefont {Ojanen}, \citenamefont {Enkovaara},\ and\ \citenamefont
  {Rantala}}]{GLLBSC}%
  \BibitemOpen
  \bibfield  {author} {\bibinfo {author} {\bibfnamefont {M.}~\bibnamefont
  {Kuisma}}, \bibinfo {author} {\bibfnamefont {J.}~\bibnamefont {Ojanen}},
  \bibinfo {author} {\bibfnamefont {J.}~\bibnamefont {Enkovaara}}, \ and\
  \bibinfo {author} {\bibfnamefont {T.~T.}\ \bibnamefont {Rantala}},\
  }\bibfield  {title} {\enquote {\bibinfo {title} {{K}ohn-{S}ham potential with
  discontinuity for band gap materials},}\ }\href {\doibase
  10.1103/PhysRevB.82.115106} {\bibfield  {journal} {\bibinfo  {journal} {Phys.
  Rev. B}\ }\textbf {\bibinfo {volume} {82}},\ \bibinfo {pages} {115106}
  (\bibinfo {year} {2010})}\BibitemShut {NoStop}%
\bibitem [{\citenamefont {Tran}\ and\ \citenamefont
  {Blaha}(2017)}]{Tran2017Importance}%
  \BibitemOpen
  \bibfield  {author} {\bibinfo {author} {\bibfnamefont {F.}~\bibnamefont
  {Tran}}\ and\ \bibinfo {author} {\bibfnamefont {P.}~\bibnamefont {Blaha}},\
  }\bibfield  {title} {\enquote {\bibinfo {title} {Importance of the kinetic
  energy density for band gap calculations in solids with density functional
  theory},}\ }\href {\doibase 10.1021/acs.jpca.7b02882} {\bibfield  {journal}
  {\bibinfo  {journal} {J. Phys. Chem. A}\ }\textbf {\bibinfo {volume} {121}},\
  \bibinfo {pages} {3318--3325} (\bibinfo {year} {2017})}\BibitemShut {NoStop}%
\bibitem [{\citenamefont {Heyd}\ \emph {et~al.}(2003)\citenamefont {Heyd},
  \citenamefont {Scuseria},\ and\ \citenamefont {Ernzerhof}}]{HSE}%
  \BibitemOpen
  \bibfield  {author} {\bibinfo {author} {\bibfnamefont {J.}~\bibnamefont
  {Heyd}}, \bibinfo {author} {\bibfnamefont {G.~E.}\ \bibnamefont {Scuseria}},
  \ and\ \bibinfo {author} {\bibfnamefont {M.}~\bibnamefont {Ernzerhof}},\
  }\bibfield  {title} {\enquote {\bibinfo {title} {Hybrid functionals based on
  a screened {C}oulomb potential},}\ }\href {\doibase 10.1063/1.1564060}
  {\bibfield  {journal} {\bibinfo  {journal} {J. Chem. Phys.}\ }\textbf
  {\bibinfo {volume} {118}},\ \bibinfo {pages} {8207} (\bibinfo {year}
  {2003})}\BibitemShut {NoStop}%
\bibitem [{\citenamefont {Onida}\ \emph {et~al.}(2002)\citenamefont {Onida},
  \citenamefont {Reining},\ and\ \citenamefont {Rubio}}]{AngelGWReview}%
  \BibitemOpen
  \bibfield  {author} {\bibinfo {author} {\bibfnamefont {G.}~\bibnamefont
  {Onida}}, \bibinfo {author} {\bibfnamefont {L.}~\bibnamefont {Reining}}, \
  and\ \bibinfo {author} {\bibfnamefont {A.}~\bibnamefont {Rubio}},\ }\bibfield
   {title} {\enquote {\bibinfo {title} {Electronic excitations:
  Density-functional versus many-body {Green}'s-function approaches},}\ }\href
  {\doibase 10.1103/RevModPhys.74.601} {\bibfield  {journal} {\bibinfo
  {journal} {Rev. Mod. Phys.}\ }\textbf {\bibinfo {volume} {74}},\ \bibinfo
  {pages} {601--659} (\bibinfo {year} {2002})}\BibitemShut {NoStop}%
\bibitem [{\citenamefont {Migani}\ \emph {et~al.}(2013)\citenamefont {Migani},
  \citenamefont {Mowbray}, \citenamefont {Iacomino}, \citenamefont {Zhao},
  \citenamefont {Petek},\ and\ \citenamefont {Rubio}}]{OurJACS}%
  \BibitemOpen
  \bibfield  {author} {\bibinfo {author} {\bibfnamefont {A.}~\bibnamefont
  {Migani}}, \bibinfo {author} {\bibfnamefont {D.~J.}\ \bibnamefont {Mowbray}},
  \bibinfo {author} {\bibfnamefont {A.}~\bibnamefont {Iacomino}}, \bibinfo
  {author} {\bibfnamefont {J.}~\bibnamefont {Zhao}}, \bibinfo {author}
  {\bibfnamefont {H.}~\bibnamefont {Petek}}, \ and\ \bibinfo {author}
  {\bibfnamefont {A.}~\bibnamefont {Rubio}},\ }\bibfield  {title} {\enquote
  {\bibinfo {title} {Level alignment of a prototypical photocatalytic system:
  Methanol on {TiO$_{\mathrm{2}}$}(110)},}\ }\href {\doibase 10.1021/ja4036994}
  {\bibfield  {journal} {\bibinfo  {journal} {J. Am. Chem. Soc.}\ }\textbf
  {\bibinfo {volume} {135}},\ \bibinfo {pages} {11429--11432} (\bibinfo {year}
  {2013})}\BibitemShut {NoStop}%
\bibitem [{\citenamefont {Tran}\ \emph {et~al.}(2018)\citenamefont {Tran},
  \citenamefont {Ehsan},\ and\ \citenamefont {Blaha}}]{Tran2018GLLBsc3}%
  \BibitemOpen
  \bibfield  {author} {\bibinfo {author} {\bibfnamefont {F.}~\bibnamefont
  {Tran}}, \bibinfo {author} {\bibfnamefont {S.}~\bibnamefont {Ehsan}}, \ and\
  \bibinfo {author} {\bibfnamefont {P.}~\bibnamefont {Blaha}},\ }\bibfield
  {title} {\enquote {\bibinfo {title} {Assessment of the {GLLB}-{SC} potential
  for solid-state properties and attempts for improvement},}\ }\href {\doibase
  10.1103/PhysRevMaterials.2.023802} {\bibfield  {journal} {\bibinfo  {journal}
  {Phys. Rev. Materials}\ }\textbf {\bibinfo {volume} {2}},\ \bibinfo {pages}
  {023802} (\bibinfo {year} {2018})}\BibitemShut {NoStop}%
\bibitem [{\citenamefont {Gritsenko}\ \emph {et~al.}(1995)\citenamefont
  {Gritsenko}, \citenamefont {van Leeuwen}, \citenamefont {van Lenthe},\ and\
  \citenamefont {Baerends}}]{Gritsenko1995GLLB}%
  \BibitemOpen
  \bibfield  {author} {\bibinfo {author} {\bibfnamefont {O.}~\bibnamefont
  {Gritsenko}}, \bibinfo {author} {\bibfnamefont {R.}~\bibnamefont {van
  Leeuwen}}, \bibinfo {author} {\bibfnamefont {E.}~\bibnamefont {van Lenthe}},
  \ and\ \bibinfo {author} {\bibfnamefont {E.~J.}\ \bibnamefont {Baerends}},\
  }\bibfield  {title} {\enquote {\bibinfo {title} {Self-consistent
  approximation to the {Kohn-Sham} exchange potential},}\ }\href {\doibase
  10.1103/physreva.51.1944} {\bibfield  {journal} {\bibinfo  {journal} {Phys.
  Rev. A}\ }\textbf {\bibinfo {volume} {51}},\ \bibinfo {pages} {1944--1954}
  (\bibinfo {year} {1995})}\BibitemShut {NoStop}%
\bibitem [{\citenamefont {Gritsenko}\ \emph {et~al.}(1997)\citenamefont
  {Gritsenko}, \citenamefont {van Leeuwen},\ and\ \citenamefont
  {Baerends}}]{Gritsenko1997GLLB2}%
  \BibitemOpen
  \bibfield  {author} {\bibinfo {author} {\bibfnamefont {O.~V.}\ \bibnamefont
  {Gritsenko}}, \bibinfo {author} {\bibfnamefont {R.}~\bibnamefont {van
  Leeuwen}}, \ and\ \bibinfo {author} {\bibfnamefont {E.~J.}\ \bibnamefont
  {Baerends}},\ }\bibfield  {title} {\enquote {\bibinfo {title} {Direct
  approximation of the long- and short-range components of the
  exchange-correlation {Kohn-Sham} potential},}\ }\href {\doibase
  10.1002/(sici)1097-461x(1997)61:2<231::aid-qua5>3.0.co;2-x} {\bibfield
  {journal} {\bibinfo  {journal} {Int. J. Quantum Chem.}\ }\textbf {\bibinfo
  {volume} {61}},\ \bibinfo {pages} {231--243} (\bibinfo {year}
  {1997})}\BibitemShut {NoStop}%
\bibitem [{\citenamefont {Castelli}\ \emph {et~al.}(2012)\citenamefont
  {Castelli}, \citenamefont {Olsen}, \citenamefont {Datta}, \citenamefont
  {Landis}, \citenamefont {Dahl}, \citenamefont {Thygesen},\ and\ \citenamefont
  {Jacobsen}}]{Castelli2012GLLBsc2}%
  \BibitemOpen
  \bibfield  {author} {\bibinfo {author} {\bibfnamefont {I.~E.}\ \bibnamefont
  {Castelli}}, \bibinfo {author} {\bibfnamefont {T.}~\bibnamefont {Olsen}},
  \bibinfo {author} {\bibfnamefont {S.}~\bibnamefont {Datta}}, \bibinfo
  {author} {\bibfnamefont {D.~D.}\ \bibnamefont {Landis}}, \bibinfo {author}
  {\bibfnamefont {S.}~\bibnamefont {Dahl}}, \bibinfo {author} {\bibfnamefont
  {K.~S.}\ \bibnamefont {Thygesen}}, \ and\ \bibinfo {author} {\bibfnamefont
  {K.~W.}\ \bibnamefont {Jacobsen}},\ }\bibfield  {title} {\enquote {\bibinfo
  {title} {Computational screening of perovskite metal oxides for optimal solar
  light capture},}\ }\href {\doibase 10.1039/c1ee02717d} {\bibfield  {journal}
  {\bibinfo  {journal} {Energy Environ. Sci.}\ }\textbf {\bibinfo {volume}
  {5}},\ \bibinfo {pages} {5814–5819} (\bibinfo {year} {2012})}\BibitemShut
  {NoStop}%
\bibitem [{\citenamefont {Glanzmann}\ \emph {et~al.}(2014)\citenamefont
  {Glanzmann}, \citenamefont {Mowbray},\ and\ \citenamefont
  {Rubio}}]{Livia2014PSSB}%
  \BibitemOpen
  \bibfield  {author} {\bibinfo {author} {\bibfnamefont {L.~N.}\ \bibnamefont
  {Glanzmann}}, \bibinfo {author} {\bibfnamefont {D.~J.}\ \bibnamefont
  {Mowbray}}, \ and\ \bibinfo {author} {\bibfnamefont {A.}~\bibnamefont
  {Rubio}},\ }\bibfield  {title} {\enquote {\bibinfo {title} {{PFO-BPy}
  solubilizers for {SWNTs}: Modelling polymers from oligomers},}\ }\href
  {\doibase 10.1002/pssb.201451171} {\bibfield  {journal} {\bibinfo  {journal}
  {Phys. Status Solidi B}\ }\textbf {\bibinfo {volume} {251}},\ \bibinfo
  {pages} {2407--2412} (\bibinfo {year} {2014})}\BibitemShut {NoStop}%
\bibitem [{\citenamefont {Mowbray}\ and\ \citenamefont
  {Migani}(2016)}]{CatecholExcitons}%
  \BibitemOpen
  \bibfield  {author} {\bibinfo {author} {\bibfnamefont {D.~J.}\ \bibnamefont
  {Mowbray}}\ and\ \bibinfo {author} {\bibfnamefont {A.}~\bibnamefont
  {Migani}},\ }\bibfield  {title} {\enquote {\bibinfo {title} {Optical
  absorption spectra and excitons of dye-substrate interfaces: Catechol on
  {TiO}$_2$(110)},}\ }\href {\doibase 10.1021/acs.jctc.6b00217} {\bibfield
  {journal} {\bibinfo  {journal} {J. Chem. Theory Comput.}\ }\textbf {\bibinfo
  {volume} {12}},\ \bibinfo {pages} {2843} (\bibinfo {year}
  {2016})}\BibitemShut {NoStop}%
\bibitem [{\citenamefont {Mortensen}\ \emph {et~al.}(2005)\citenamefont
  {Mortensen}, \citenamefont {Hansen},\ and\ \citenamefont {Jacobsen}}]{GPAW}%
  \BibitemOpen
  \bibfield  {author} {\bibinfo {author} {\bibfnamefont {J.~J.}\ \bibnamefont
  {Mortensen}}, \bibinfo {author} {\bibfnamefont {L.~B.}\ \bibnamefont
  {Hansen}}, \ and\ \bibinfo {author} {\bibfnamefont {K.~W.}\ \bibnamefont
  {Jacobsen}},\ }\bibfield  {title} {\enquote {\bibinfo {title} {Real-space
  grid implementation of the projector augmented wave method},}\ }\href
  {\doibase 10.1103/PhysRevB.71.035109} {\bibfield  {journal} {\bibinfo
  {journal} {Phys. Rev. B}\ }\textbf {\bibinfo {volume} {71}},\ \bibinfo
  {pages} {035109} (\bibinfo {year} {2005})}\BibitemShut {NoStop}%
\bibitem [{\citenamefont {Enkovaara}\ \emph {et~al.}(2010)\citenamefont
  {Enkovaara}, \citenamefont {Rostgaard}, \citenamefont {Mortensen},
  \citenamefont {Chen}, \citenamefont {Du{\l}ak}, \citenamefont {Ferrighi},
  \citenamefont {Gavnholt}, \citenamefont {Glinsvad}, \citenamefont {Haikola},
  \citenamefont {Hansen}, \citenamefont {Kristoffersen}, \citenamefont
  {Kuisma}, \citenamefont {Larsen}, \citenamefont {Lehtovaara}, \citenamefont
  {Ljungberg}, \citenamefont {Lopez-Acevedo}, \citenamefont {Moses},
  \citenamefont {Ojanen}, \citenamefont {Olsen}, \citenamefont {Petzold},
  \citenamefont {Romero}, \citenamefont {Stausholm-M{\o}ller}, \citenamefont
  {Strange}, \citenamefont {Tritsaris}, \citenamefont {Vanin}, \citenamefont
  {Walter}, \citenamefont {Hammer}, \citenamefont {H{\"{a}}kkinen},
  \citenamefont {Madsen}, \citenamefont {Nieminen}, \citenamefont {N{\o}rskov},
  \citenamefont {Puska}, \citenamefont {Rantala}, \citenamefont {Schi{\o}tz},
  \citenamefont {Thygesen},\ and\ \citenamefont {Jacobsen}}]{GPAWrev}%
  \BibitemOpen
  \bibfield  {author} {\bibinfo {author} {\bibfnamefont {J.}~\bibnamefont
  {Enkovaara}}, \bibinfo {author} {\bibfnamefont {C.}~\bibnamefont
  {Rostgaard}}, \bibinfo {author} {\bibfnamefont {J.~J.}\ \bibnamefont
  {Mortensen}}, \bibinfo {author} {\bibfnamefont {J.}~\bibnamefont {Chen}},
  \bibinfo {author} {\bibfnamefont {M.}~\bibnamefont {Du{\l}ak}}, \bibinfo
  {author} {\bibfnamefont {L.}~\bibnamefont {Ferrighi}}, \bibinfo {author}
  {\bibfnamefont {J.}~\bibnamefont {Gavnholt}}, \bibinfo {author}
  {\bibfnamefont {C.}~\bibnamefont {Glinsvad}}, \bibinfo {author}
  {\bibfnamefont {V.}~\bibnamefont {Haikola}}, \bibinfo {author} {\bibfnamefont
  {H.~A.}\ \bibnamefont {Hansen}}, \bibinfo {author} {\bibfnamefont {H.~H.}\
  \bibnamefont {Kristoffersen}}, \bibinfo {author} {\bibfnamefont
  {M.}~\bibnamefont {Kuisma}}, \bibinfo {author} {\bibfnamefont {A.~H.}\
  \bibnamefont {Larsen}}, \bibinfo {author} {\bibfnamefont {L.}~\bibnamefont
  {Lehtovaara}}, \bibinfo {author} {\bibfnamefont {M.}~\bibnamefont
  {Ljungberg}}, \bibinfo {author} {\bibfnamefont {O.}~\bibnamefont
  {Lopez-Acevedo}}, \bibinfo {author} {\bibfnamefont {P.~G.}\ \bibnamefont
  {Moses}}, \bibinfo {author} {\bibfnamefont {J.}~\bibnamefont {Ojanen}},
  \bibinfo {author} {\bibfnamefont {T.}~\bibnamefont {Olsen}}, \bibinfo
  {author} {\bibfnamefont {V.}~\bibnamefont {Petzold}}, \bibinfo {author}
  {\bibfnamefont {N.~A.}\ \bibnamefont {Romero}}, \bibinfo {author}
  {\bibfnamefont {J.}~\bibnamefont {Stausholm-M{\o}ller}}, \bibinfo {author}
  {\bibfnamefont {M.}~\bibnamefont {Strange}}, \bibinfo {author} {\bibfnamefont
  {G.~A.}\ \bibnamefont {Tritsaris}}, \bibinfo {author} {\bibfnamefont
  {M.}~\bibnamefont {Vanin}}, \bibinfo {author} {\bibfnamefont
  {M.}~\bibnamefont {Walter}}, \bibinfo {author} {\bibfnamefont
  {B.}~\bibnamefont {Hammer}}, \bibinfo {author} {\bibfnamefont
  {H.}~\bibnamefont {H{\"{a}}kkinen}}, \bibinfo {author} {\bibfnamefont
  {G.~K.~H.}\ \bibnamefont {Madsen}}, \bibinfo {author} {\bibfnamefont {R.~M.}\
  \bibnamefont {Nieminen}}, \bibinfo {author} {\bibfnamefont {J.~K.}\
  \bibnamefont {N{\o}rskov}}, \bibinfo {author} {\bibfnamefont
  {M.}~\bibnamefont {Puska}}, \bibinfo {author} {\bibfnamefont {T.~T.}\
  \bibnamefont {Rantala}}, \bibinfo {author} {\bibfnamefont {J.}~\bibnamefont
  {Schi{\o}tz}}, \bibinfo {author} {\bibfnamefont {K.~S.}\ \bibnamefont
  {Thygesen}}, \ and\ \bibinfo {author} {\bibfnamefont {K.~W.}\ \bibnamefont
  {Jacobsen}},\ }\bibfield  {title} {\enquote {\bibinfo {title} {Electronic
  structure calculations with {GPAW}: A real-space implementation of the
  projector augmented-wave method},}\ }\href {\doibase
  10.1088/0953-8984/22/25/253202} {\bibfield  {journal} {\bibinfo  {journal}
  {J. Phys.: Condens. Matter}\ }\textbf {\bibinfo {volume} {22}},\ \bibinfo
  {pages} {253202} (\bibinfo {year} {2010})}\BibitemShut {NoStop}%
\bibitem [{\citenamefont {Bl{\"{o}}chl}(1994)}]{PAW}%
  \BibitemOpen
  \bibfield  {author} {\bibinfo {author} {\bibfnamefont {P.~E.}\ \bibnamefont
  {Bl{\"{o}}chl}},\ }\bibfield  {title} {\enquote {\bibinfo {title} {Projector
  augmented-wave method},}\ }\href {\doibase 10.1103/PhysRevB.50.17953}
  {\bibfield  {journal} {\bibinfo  {journal} {Phys. Rev. B}\ }\textbf {\bibinfo
  {volume} {50}},\ \bibinfo {pages} {17953--17979} (\bibinfo {year}
  {1994})}\BibitemShut {NoStop}%
\bibitem [{\citenamefont {Bahn}\ and\ \citenamefont {Jacobsen}(2002)}]{ASE0}%
  \BibitemOpen
  \bibfield  {author} {\bibinfo {author} {\bibfnamefont {S.~R.}\ \bibnamefont
  {Bahn}}\ and\ \bibinfo {author} {\bibfnamefont {K.~W.}\ \bibnamefont
  {Jacobsen}},\ }\bibfield  {title} {\enquote {\bibinfo {title} {An
  object-oriented scripting interface to a legacy electronic structure code},}\
  }\href {\doibase 10.1109/5992.998641} {\bibfield  {journal} {\bibinfo
  {journal} {Comput. Sci. Eng.}\ }\textbf {\bibinfo {volume} {4}},\ \bibinfo
  {pages} {56--66} (\bibinfo {year} {2002})}\BibitemShut {NoStop}%
\bibitem [{\citenamefont {Larsen}\ \emph {et~al.}(2017)\citenamefont {Larsen},
  \citenamefont {Mortensen}, \citenamefont {Blomqvist}, \citenamefont
  {Castelli}, \citenamefont {Christensen}, \citenamefont {Du{\l}ak},
  \citenamefont {Friis}, \citenamefont {Groves}, \citenamefont {Hammer},
  \citenamefont {Hargus}, \citenamefont {Hermes}, \citenamefont {Jennings},
  \citenamefont {Jensen}, \citenamefont {Kermode}, \citenamefont {Kitchin},
  \citenamefont {Kolsbjerg}, \citenamefont {Kubal}, \citenamefont {Kaasbjerg},
  \citenamefont {Lysgaard}, \citenamefont {Maronsson}, \citenamefont {Maxson},
  \citenamefont {Olsen}, \citenamefont {Pastewka}, \citenamefont {Peterson},
  \citenamefont {Rostgaard}, \citenamefont {Schi{\o}tz}, \citenamefont
  {Sch{\"{u}}tt}, \citenamefont {Strange}, \citenamefont {Thygesen},
  \citenamefont {Vegge}, \citenamefont {Vilhelmsen}, \citenamefont {Walter},
  \citenamefont {Zeng},\ and\ \citenamefont {Jacobsen}}]{ASE}%
  \BibitemOpen
  \bibfield  {author} {\bibinfo {author} {\bibfnamefont {A.~H.}\ \bibnamefont
  {Larsen}}, \bibinfo {author} {\bibfnamefont {J.~J.}\ \bibnamefont
  {Mortensen}}, \bibinfo {author} {\bibfnamefont {J.}~\bibnamefont
  {Blomqvist}}, \bibinfo {author} {\bibfnamefont {I.~E.}\ \bibnamefont
  {Castelli}}, \bibinfo {author} {\bibfnamefont {R.}~\bibnamefont
  {Christensen}}, \bibinfo {author} {\bibfnamefont {M.}~\bibnamefont
  {Du{\l}ak}}, \bibinfo {author} {\bibfnamefont {J.}~\bibnamefont {Friis}},
  \bibinfo {author} {\bibfnamefont {M.~N.}\ \bibnamefont {Groves}}, \bibinfo
  {author} {\bibfnamefont {B.}~\bibnamefont {Hammer}}, \bibinfo {author}
  {\bibfnamefont {C.}~\bibnamefont {Hargus}}, \bibinfo {author} {\bibfnamefont
  {E.~D.}\ \bibnamefont {Hermes}}, \bibinfo {author} {\bibfnamefont {P.~C.}\
  \bibnamefont {Jennings}}, \bibinfo {author} {\bibfnamefont {P.~B.}\
  \bibnamefont {Jensen}}, \bibinfo {author} {\bibfnamefont {J.}~\bibnamefont
  {Kermode}}, \bibinfo {author} {\bibfnamefont {J.~R.}\ \bibnamefont
  {Kitchin}}, \bibinfo {author} {\bibfnamefont {E.~L.}\ \bibnamefont
  {Kolsbjerg}}, \bibinfo {author} {\bibfnamefont {J.}~\bibnamefont {Kubal}},
  \bibinfo {author} {\bibfnamefont {K.}~\bibnamefont {Kaasbjerg}}, \bibinfo
  {author} {\bibfnamefont {S.}~\bibnamefont {Lysgaard}}, \bibinfo {author}
  {\bibfnamefont {J.~B.}\ \bibnamefont {Maronsson}}, \bibinfo {author}
  {\bibfnamefont {T.}~\bibnamefont {Maxson}}, \bibinfo {author} {\bibfnamefont
  {T.}~\bibnamefont {Olsen}}, \bibinfo {author} {\bibfnamefont
  {L.}~\bibnamefont {Pastewka}}, \bibinfo {author} {\bibfnamefont
  {A.}~\bibnamefont {Peterson}}, \bibinfo {author} {\bibfnamefont
  {C.}~\bibnamefont {Rostgaard}}, \bibinfo {author} {\bibfnamefont
  {J.}~\bibnamefont {Schi{\o}tz}}, \bibinfo {author} {\bibfnamefont
  {O.}~\bibnamefont {Sch{\"{u}}tt}}, \bibinfo {author} {\bibfnamefont
  {M.}~\bibnamefont {Strange}}, \bibinfo {author} {\bibfnamefont {K.~S.}\
  \bibnamefont {Thygesen}}, \bibinfo {author} {\bibfnamefont {T.}~\bibnamefont
  {Vegge}}, \bibinfo {author} {\bibfnamefont {L.}~\bibnamefont {Vilhelmsen}},
  \bibinfo {author} {\bibfnamefont {M.}~\bibnamefont {Walter}}, \bibinfo
  {author} {\bibfnamefont {Z.}~\bibnamefont {Zeng}}, \ and\ \bibinfo {author}
  {\bibfnamefont {K.~W.}\ \bibnamefont {Jacobsen}},\ }\bibfield  {title}
  {\enquote {\bibinfo {title} {The atomic simulation environment---a python
  library for working with atoms},}\ }\href {\doibase 10.1088/1361-648X/aa680e}
  {\bibfield  {journal} {\bibinfo  {journal} {J. Phys.: Condens. Matter}\
  }\textbf {\bibinfo {volume} {29}},\ \bibinfo {pages} {273002} (\bibinfo
  {year} {2017})}\BibitemShut {NoStop}%
\bibitem [{\citenamefont {Perdew}\ \emph {et~al.}(2008)\citenamefont {Perdew},
  \citenamefont {Ruzsinszky}, \citenamefont {Csonka}, \citenamefont {Vydrov},
  \citenamefont {Scuseria}, \citenamefont {Constantin}, \citenamefont {Zhou},\
  and\ \citenamefont {Burke}}]{PBEsol}%
  \BibitemOpen
  \bibfield  {author} {\bibinfo {author} {\bibfnamefont {J.~P.}\ \bibnamefont
  {Perdew}}, \bibinfo {author} {\bibfnamefont {A.}~\bibnamefont {Ruzsinszky}},
  \bibinfo {author} {\bibfnamefont {G.~I.}\ \bibnamefont {Csonka}}, \bibinfo
  {author} {\bibfnamefont {O.~A.}\ \bibnamefont {Vydrov}}, \bibinfo {author}
  {\bibfnamefont {G.~E.}\ \bibnamefont {Scuseria}}, \bibinfo {author}
  {\bibfnamefont {L.~A.}\ \bibnamefont {Constantin}}, \bibinfo {author}
  {\bibfnamefont {X.}~\bibnamefont {Zhou}}, \ and\ \bibinfo {author}
  {\bibfnamefont {K.}~\bibnamefont {Burke}},\ }\bibfield  {title} {\enquote
  {\bibinfo {title} {Restoring the density-gradient expansion for exchange in
  solids and surfaces},}\ }\href {\doibase 10.1103/PhysRevLett.100.136406}
  {\bibfield  {journal} {\bibinfo  {journal} {Phys. Rev. Lett.}\ }\textbf
  {\bibinfo {volume} {100}},\ \bibinfo {pages} {136406} (\bibinfo {year}
  {2008})}\BibitemShut {NoStop}%
\bibitem [{\citenamefont {Larsen}\ \emph {et~al.}(2009)\citenamefont {Larsen},
  \citenamefont {Vanin}, \citenamefont {Mortensen}, \citenamefont {Thygesen},\
  and\ \citenamefont {Jacobsen}}]{GPAWLCAO}%
  \BibitemOpen
  \bibfield  {author} {\bibinfo {author} {\bibfnamefont {A.~H.}\ \bibnamefont
  {Larsen}}, \bibinfo {author} {\bibfnamefont {M.}~\bibnamefont {Vanin}},
  \bibinfo {author} {\bibfnamefont {J.~J.}\ \bibnamefont {Mortensen}}, \bibinfo
  {author} {\bibfnamefont {K.~S.}\ \bibnamefont {Thygesen}}, \ and\ \bibinfo
  {author} {\bibfnamefont {K.~W.}\ \bibnamefont {Jacobsen}},\ }\bibfield
  {title} {\enquote {\bibinfo {title} {Localized atomic basis set in the
  projector augmented wave method},}\ }\href {\doibase
  10.1103/PhysRevB.80.195112} {\bibfield  {journal} {\bibinfo  {journal} {Phys.
  Rev. B}\ }\textbf {\bibinfo {volume} {80}},\ \bibinfo {pages} {195112}
  (\bibinfo {year} {2009})}\BibitemShut {NoStop}%
\bibitem [{\citenamefont {Pichler}\ \emph {et~al.}(1998)\citenamefont
  {Pichler}, \citenamefont {Knupfer}, \citenamefont {Golden}, \citenamefont
  {Fink}, \citenamefont {Rinzler},\ and\ \citenamefont
  {Smalley}}]{SWCNTpiEELS99}%
  \BibitemOpen
  \bibfield  {author} {\bibinfo {author} {\bibfnamefont {T.}~\bibnamefont
  {Pichler}}, \bibinfo {author} {\bibfnamefont {M.}~\bibnamefont {Knupfer}},
  \bibinfo {author} {\bibfnamefont {M.~S.}\ \bibnamefont {Golden}}, \bibinfo
  {author} {\bibfnamefont {J.}~\bibnamefont {Fink}}, \bibinfo {author}
  {\bibfnamefont {A.}~\bibnamefont {Rinzler}}, \ and\ \bibinfo {author}
  {\bibfnamefont {R.~E.}\ \bibnamefont {Smalley}},\ }\bibfield  {title}
  {\enquote {\bibinfo {title} {Localized and delocalized electronic states in
  single-wall carbon nanotubes},}\ }\href {\doibase
  10.1103/PhysRevLett.80.4729} {\bibfield  {journal} {\bibinfo  {journal}
  {Phys. Rev. Lett.}\ }\textbf {\bibinfo {volume} {80}},\ \bibinfo {pages}
  {4729--4732} (\bibinfo {year} {1998})}\BibitemShut {NoStop}%
\bibitem [{\citenamefont {Kramberger}\ \emph {et~al.}(2008)\citenamefont
  {Kramberger}, \citenamefont {Hambach}, \citenamefont {Giorgetti},
  \citenamefont {R\"ummeli}, \citenamefont {Knupfer}, \citenamefont {Fink},
  \citenamefont {B\"uchner}, \citenamefont {Reining}, \citenamefont
  {Einarsson}, \citenamefont {Maruyama}, \citenamefont {Sottile}, \citenamefont
  {Hannewald}, \citenamefont {Olevano}, \citenamefont {Marinopoulos},\ and\
  \citenamefont {Pichler}}]{SWCNTpiEELS08}%
  \BibitemOpen
  \bibfield  {author} {\bibinfo {author} {\bibfnamefont {C.}~\bibnamefont
  {Kramberger}}, \bibinfo {author} {\bibfnamefont {R.}~\bibnamefont {Hambach}},
  \bibinfo {author} {\bibfnamefont {C.}~\bibnamefont {Giorgetti}}, \bibinfo
  {author} {\bibfnamefont {M.~H.}\ \bibnamefont {R\"ummeli}}, \bibinfo {author}
  {\bibfnamefont {M.}~\bibnamefont {Knupfer}}, \bibinfo {author} {\bibfnamefont
  {J.}~\bibnamefont {Fink}}, \bibinfo {author} {\bibfnamefont {B.}~\bibnamefont
  {B\"uchner}}, \bibinfo {author} {\bibfnamefont {L.}~\bibnamefont {Reining}},
  \bibinfo {author} {\bibfnamefont {E.}~\bibnamefont {Einarsson}}, \bibinfo
  {author} {\bibfnamefont {S.}~\bibnamefont {Maruyama}}, \bibinfo {author}
  {\bibfnamefont {F.}~\bibnamefont {Sottile}}, \bibinfo {author} {\bibfnamefont
  {K.}~\bibnamefont {Hannewald}}, \bibinfo {author} {\bibfnamefont
  {V.}~\bibnamefont {Olevano}}, \bibinfo {author} {\bibfnamefont {A.~G.}\
  \bibnamefont {Marinopoulos}}, \ and\ \bibinfo {author} {\bibfnamefont
  {T.}~\bibnamefont {Pichler}},\ }\bibfield  {title} {\enquote {\bibinfo
  {title} {Linear plasmon dispersion in single-wall carbon nanotubes and the
  collective excitation spectrum of graphene},}\ }\href {\doibase
  10.1103/PhysRevLett.100.196803} {\bibfield  {journal} {\bibinfo  {journal}
  {Phys. Rev. Lett.}\ }\textbf {\bibinfo {volume} {100}},\ \bibinfo {pages}
  {196803} (\bibinfo {year} {2008})}\BibitemShut {NoStop}%
\bibitem [{\citenamefont {Mowbray}\ \emph {et~al.}(2010)\citenamefont
  {Mowbray}, \citenamefont {Segui}, \citenamefont {Gervasoni}, \citenamefont
  {Mi{\v{s}}kovi{\'{c}}},\ and\ \citenamefont {Arista}}]{Silvina}%
  \BibitemOpen
  \bibfield  {author} {\bibinfo {author} {\bibfnamefont {D.~J.}\ \bibnamefont
  {Mowbray}}, \bibinfo {author} {\bibfnamefont {S.}~\bibnamefont {Segui}},
  \bibinfo {author} {\bibfnamefont {J.}~\bibnamefont {Gervasoni}}, \bibinfo
  {author} {\bibfnamefont {Z.~L.}\ \bibnamefont {Mi{\v{s}}kovi{\'{c}}}}, \ and\
  \bibinfo {author} {\bibfnamefont {N.~R.}\ \bibnamefont {Arista}},\ }\bibfield
   {title} {\enquote {\bibinfo {title} {Plasmon excitations on a single-wall
  carbon nanotube by external charges: Two-dimensional, two-fluid hydrodynamic
  model},}\ }\href {\doibase 10.1103/PhysRevB.82.035405} {\bibfield  {journal}
  {\bibinfo  {journal} {Phys. Rev. B}\ }\textbf {\bibinfo {volume} {82}},\
  \bibinfo {pages} {035405} (\bibinfo {year} {2010})}\BibitemShut {NoStop}%
\end{thebibliography}
\end{document}